\documentclass{emulateapj}
\usepackage{amsmath}

\begin{document}

\title{The Impact of Environment on the Stellar Mass -- Halo Mass Relation}
\author{Jesse B. Golden-Marx \altaffilmark{1} \and  Christopher J. Miller\altaffilmark{1,2,3}}
\altaffiltext{1}{Department of Astronomy, University of Michigan, Ann Arbor, MI 48109 USA}
\altaffiltext{2}{Department of Physics, University of Michigan, Ann Arbor, MI 48109, USA}
\altaffiltext{3}{Asa Briggs Fellow, University of Sussex, Brighton, BN1 9RH,
United Kingdom}
\email{jessegm@umich.edu}

\slugcomment{Received 2017 October 30, Accepted to ApJ, 2018 May 2}
\begin{abstract}

A large variance exists in the amplitude of the Stellar Mass -- Halo Mass (SMHM) relation for group and cluster-size halos.  Using a sample of 254 clusters, we show that the magnitude gap between the brightest central galaxy (BCG) and its second or fourth brightest neighbor accounts for a significant portion of this variance.  We find that at fixed halo mass, galaxy clusters with a higher magnitude gap have a higher BCG stellar mass.  This relationship is also observed in semi-analytic representations of low-redshift galaxy clusters in simulations.  This SMHM-magnitude gap stratification likely results from BCG growth via hierarchical mergers and may link assembly of the halo with the growth of the BCG.  Using a Bayesian model, we quantify the importance of the magnitude gap in the SMHM relation using a multiplicative stretch factor, which we find to be significantly non-zero.  The inclusion of the magnitude gap in the SMHM relation results in a large reduction in the inferred intrinsic scatter in the BCG stellar mass at fixed halo mass.  We discuss the ramifications of this result in the context of galaxy formation models of centrals in group and cluster-sized halos.
\end{abstract}

\section{Introduction} \label{sec:intro}
At the heart of galaxy clusters lie brightest central galaxies, or BCGs.  These galaxies are bright, often 10$L_{\star}$, where $L_{\star}$ is the characteristic luminosity of the galaxy luminosity function \citep[e.g.,][]{sch86}, and extremely massive as shown by dynamical mass estimates \citep[e.g.,][]{ber07, von07,bro11,pro11}.  Studies have also shown that BCGs can account for a substantial fraction of the total light emitted from a galaxy cluster \citep[e.g.,][]{jon00, lin04, agu11, har12}.  Additionally, BCGs are often found near the X-ray centers of galaxy clusters \citep[e.g.,][]{jon84, rhe91, lin04, lauer14}.  These massive central galaxies can also be more spatially extended than similarly massive elliptical galaxies \citep{ber07, lau07} and are often surrounded by halos of diffuse intracluster light \citep[e.g.,][]{zwi51, wel71, oem76, lin04}.

We note that not all central galaxies match the standard definition of a BCG.  \cite{skibba11} and \citet{lan18} suggest that as many as 40\% of massive, low redshift clusters have the equivalent of a satellite galaxy as their central.  In contrast, \citet{lauer14} find that only 15\% of all BCGs in their low redshift sample have a separation between the x-ray center and BCG greater than 100 kpc.  While these results differ, it is likely that some fraction of BCGs are not always located at the cluster center-of-mass.     

Researchers have been studying BCGs to understand their growth history for over forty years \citep[e.g.,][]{ost75,ost77,hau78,mal84,mer85,fab94,ara98,dub98,del07,rus09,lid12,lap13,lin13,lauer14,nipoti17}.  Under the hierarchical structure formation paradigm, we might naturally expect some trends in the observable properties of BCGs to be caused by growth mechanisms that are not characteristic of the wider galaxy population.  For instance, \cite{lauer14} conclude that the extended envelopes present in many BCGs are formed by processes within the cluster core region.  They also suggest that especially bright BCG luminosities stem from accretion into more massive clusters. 

In practice, a property used to differentiate BCGs from non-BCGs is the ``magnitude gap,'' a measure of the difference in brightness between the BCG and some lesser galaxy within the cluster.  Early on, it was suggested that BCGs and their magnitude gaps evolve through a process that differs from normal galaxies \citep[e.g.,][]{tre77, sch83, bha85, pos95, bernstein01, vale08}.  However, some recent studies, which use large cluster samples, argue that BCGs are simply statistical draws from the extreme bright end of the galaxy luminosity function \citep[e.g.,][]{lin10, paranjape12}.  While this debate is not entirely settled, there exists a growing consensus that at least some component of the BCG population is distinct from the nominal distribution of elliptical galaxies \citep[e.g.,][]{loh06, vale08, collins09, lin10, lauer14, shen14, zhang16}.

From a theoretical perspective, \cite{milo05} showed that excursion-set merger probabilities and the standard theory of dynamical segregation can explain the distribution of BCG magnitude gaps in low redshift clusters.  At the same time, state-of-the-art cosmological simulations with semi-analytic and semi-empirical prescriptions for the growth of the stellar properties of galaxies also support the observational consensus of standard hierarchical mechanisms as the dominant influence on the growth of BCGs \citep[e.g.,][]{cro06, del07, ton12, shankar15}.  In other words, both the theory and data are converging onto a scenario that links the growth of BCGs to the earliest formation environments of their host halos.

Recently, \cite{solanes16} used dissipationless simulations of young and pre-virialized groups to show that the magnitude gap between the BCG and second brightest cluster galaxy correlates with the initial stellar mass fraction of the parent cluster halo.  This correlation suggests that the observed magnitude gap can inform us about the underlying normal mass (both stellar and baryonic) of a cluster.  For example, by identifying clusters with large magnitude gaps, we may simultaneously be identifying clusters with high stellar mass fractions during the epoch of BCG formation.  Additionally, \citet{solanes16} found that the magnitude gap contains information about the BCG's merger history.  In agreement with hierarchical growth, they found that a BCG's stellar mass increases with the number of progenitor galaxies (i.e., the number of mergers).  Moreover, \citet{solanes16} found that BCGs grow at the expense of the second brightest galaxy in the cluster.  Thus, as the BCG brightens, the cluster member identified as the 2nd brightest galaxy becomes fainter, relative to the BCG, leading to an increase in the magnitude gap.  Combining these results implies that the magnitude gap not only correlates with the stellar mass of the BCG, but also provides information about the BCG's merger history.  

One way to extend our understanding of how BCG properties relate to the host halo is to utilize the observed stellar mass -- halo mass (SMHM) relation for clusters, which directly compares the amount of stellar mass within the central galaxy of the halo (i.e., the BCG) to the overall halo mass, including the baryonic and dark matter within the cluster.  One of the earlier cluster-scale SMHM relations, presented in Figure 3 of \citet{lin04}, illustrates that the BCG luminosity, which relates to the BCG stellar mass, linearly correlates with the halo mass.  Since the work of \cite{lin04}, there have been many characterizations of the SMHM relation \citep[e.g.,][]{yan09,mos10,beh10,beh13,mos13,tin16,kra14} across a much larger range in halo mass.

When one compares the high-mass end of the SMHM relation from \citet{lin04}, \citet{han09}, \citet{beh10}, \citet{mos10}, \citet{beh13} \citet{ber13}, \citet{mos13}, \citet{tin16}, and \citet{kra14} there are differences in the inferred amplitude as large as an order of magnitude in stellar mass at fixed halo mass (see Figure \ref{fig:SMHMR-C4-Comparison}).  One challenge when comparing published cluster-scale SMHM relations is that they use different cluster/BCG samples with different selection criteria.  Additionally, there are differences in how BCG stellar masses are inferred (i.e., different Initial Mass Functions (IMFs), Stellar Population Synthesis (SPS) models, and Star Formation Histories), and how cluster (halo) masses are measured.  There can even be differences in how the BCG magnitudes are measured \citep[e.g.,][]{kra14}. A fair comparison between the previously published cluster-scale SMHM relations has yet to be reported. 

\citet{har12} presented a cluster-scale SMHM relation and compared a sample of high magnitude gap X-ray clusters, fossil galaxies, to a normal-magnitude gap cluster population.  In both cases, clusters were X-ray selected to minimize selection variations, and the BCG stellar masses were inferred using the same model.  Instead of halo masses, cluster X-ray temperature was used as a halo mass proxy.  In other words, \citet{har12} compared the SMHM relation for two cluster samples whose only difference was the magnitude gap.  They found that for a given halo mass, galaxy clusters with larger magnitude gaps have BCGs with higher stellar masses than clusters with smaller magnitude gaps.  This bifurcation between large-gap and small-gap clusters has also been previously observed for both cluster and group size halos in both simulations \citep{dia08,kun17} and in other observed samples \citep{zar14,tre16}.  In addition, the \citet{har12} results suggest that perhaps as much as half of the scatter in BCG stellar mass at fixed halo mass may be accounted for by the magnitude gap.  These previous results and studies lead us to explore the possibility that the SMHM relation contains the magnitude gap as a latent variable which, if properly accounted for, could reduce the intrinsic scatter in the stellar mass at fixed halo mass and simultaneously inform us about the formation history of both the BCG and the parent halo.

The outline for the remainder of the paper is as follows.  In Sections~\ref{sec:data} and \ref{sec:Sims} we describe the observational and simulated data used to determine the stellar masses, halo masses, and magnitude gaps that are used in our analysis of the SMHM relation.  In Section~\ref{sec:model} we describe the Bayesian MCMC model used to evaluate the SMHM relation.  In Section~\ref{sec:Results} we present the results of our analysis for both the observations and simulations, which includes a quantitative measure on the impact of incorporating the magnitude gap.  Lastly, we discuss our results in the context of galaxy formation scenarios in Section~\ref{sec:Discussion}.

Except for the case of the simulated data, in which the cosmological parameters are previously defined \citep{spr05}, for our analysis, we assume a flat $\Lambda$CDM universe, with $\Omega_{M}$=0.30, $\Omega_{\Lambda}$=0.70, H=100~$h$~km/s/Mpc with $h$=0.7.

\section{The Data} \label{sec:data}
We use data from the Sloan Digital Sky Survey data release 12 \citep[SDSS DR12;][]{alam15} to identify the clusters, measure the cluster masses, identify the BCGs, characterize their magnitude gaps, and estimate their stellar masses.  We discuss each of these in the following subsections.

\subsection{The SDSS-C4 Clusters and Dynamical Masses}
\label{subsec:SDSS-C4}
We use galaxy clusters identified using the C4 algorithm \citep{mil05} on the SDSS DR12 data \citep{alam15}. The algorithm identifies 970 clusters between $0.03 \le z \leq 0.18$.  As described in detail in \cite{mil05}, the C4 algorithm uses the four colors from the SDSS galaxy main sample and applies a non-parametric algorithm to identify statistical over-densities in color and position space.  As in \cite{mil05}, only galaxies with spectra are used to identify candidate clusters, the larger SDSS photometric sample is then used to quantify the BCG magnitudes and the magnitude gap.

We then use the spectroscopically confirmed clusters to create radius/velocity phase-spaces of the galaxies projected along the line-of-sight to the clusters and relative to the mean velocity of the cluster members.  We calculate ``caustic'' masses according to the algorithm defined in \cite{gif13a} for each cluster.  Specifically, we identify the phase-space edge as a proxy for the projected escape velocity profile of each cluster individually.  To infer masses from the projected escape velocity, the ``caustic'' technique requires a calibration term based on the unknown velocity anisotropy $\beta$.  This term is typically referred to as $\mathcal{F}_{\beta}$ and we choose a value that calibrates ``caustic'' masses in N-body simulations, $\mathcal{F}_{\beta}$ = 0.65 \citep{gif13a}.  The uncertainty on $\mathcal{F}_{\beta}$ results from the projection of the three-dimensional velocities along the line-of-sight and is the dominant component of the error on the caustic cluster masses. \cite{gif13a} used simulations to also calculate the scatter in the true mass versus the ``caustic'' mass as a function of the number of galaxies used to construct the phase-spaces.  This scatter, caused by the line-of-sight projection is the dominant component of the error in the caustic masses \citep{gif13b,gif17}.  The mass error can be large for poorly sampled phase-spaces (e.g., 0.9 dex when $N_{gal}$=10) and has a floor of about 0.3 dex for well-sampled phase-spaces ($N_{gal} >$ 150).  Therefore, we use heteroskedastic cluster mass errors based on Table 1 of \cite{gif13a}. 

\subsection{Candidate Cluster Sample}
\label{subsec:Cluster-Sample}
To construct the cleanest SMHM relation for our C4 clusters, we applied additional cuts to the cluster sample.  These cuts are summarized in Table~\ref{tab:SDSS-C4-data}.  We require $log(M_{caustic}/h) \ge 14.0$ to ensure higher completeness of our sample.  This cut reduces the total number of clusters from 970 to 420.  We then further reduced our sample by analyzing the caustic phase space, velocity histogram, and red sequence within $R_{vir}$ from the cluster center for each cluster.  We removed 32 clusters which had either a broad and unpeaked velocity histogram, indistinguishable red sequence within $R_{vir}$, or poorly defined caustic phase space, leaving us with a sample of 388 clusters.  The number of galaxies used to construct the cluster phase-spaces ranged from 31 to 1074 with a median of 122 for the remaining clusters.  Additionally, 6 more clusters were removed due to photometric issues in the SDSS pipeline, leaving us with 382 clusters.  
 
For the purpose of measuring the magnitude gap between the BCG and 2nd or 4th brightest galaxies, we also require that each cluster contain 4 members identified using the red sequence within 0.5 $R_{vir}$.\footnote{We use 0.5 $R_{vir}$ to determine these magnitude gaps because it is the standard radius used in the definitions of fossil galaxies presented in \citet{jon03} and \citet{dar10}.}  Our process of identifying cluster members will be described in further depth in Section~\ref{subsec:Maggap}.  This cut reduced our final sample by 12 clusters, leaving us with an initial sample of 370 clusters. Additional cuts were made based solely on BCG photometry, which we describe below.    

\subsection{BCG Identification and Characterization}
\label{subsec:BCG-ID}
For the majority of clusters the BCG is clearly the brightest galaxy and can be easily identified algorithmically using the red-sequence; however, not all clusters have a clear and distinctive BCG \citep{von07,lauer14}. Therefore, we visually inspect images of each cluster in regions out to 0.5 $R_{vir}$ radial.  For approximately 70\% of our galaxy clusters, the visual checks confirm the BCGs through a simple selection algorithm (e.g., \cite{von07}). In the other cases, the BCGs are identified after allowing for positional offsets from the originally defined cluster centers (still within 0.5 Mpc$h^{-1}$), or by allowing for BCG colors that are bluer than the red sequence (by up to 0.2 magnitudes in $g-r$ color).  In these cases, we selected elliptical BCGs that had positions, redshifts, and colors which matched the cluster and red sequence.  In a few ($<$2.5\%) cases, it was not clear which of the two brightest galaxies was the BCG, so we chose at random.  Since the brightness and color of these galaxies are similar, this choice will make no discernible difference in the measurements of either the stellar mass, magnitude gap, or cluster (halo) mass.

\subsubsection{BCG Stellar Mass Measurements}
\label{subsubsec:BCG-Mass}
To estimate BCG stellar masses, we require accurate extinction corrected BCG apparent magnitudes.  The SDSS DR12 photometric pipeline overestimates the light contribution from the local background for large, extended objects, and crowded fields.  Therefore, BCGs are likely to be affected by this background overestimation, which would lead to an under-approximation of the stellar mass \citep[e.g.,][]{ber07, von07, har12, ber13}.

We correct the BCG Petrosian magnitudes using the prescription outlined in the appendix of \citet{von07} applied to the SDSS DR12 apparent g, r, and i-band magnitudes of our BCGs.  The \citet{von07} correction works by adding a fraction of the difference between the SDSS local and global sky backgrounds to each individual galaxy's light profile.  Since the original \citet{von07} correction was calibrated using SDSS DR4 data, we re-calibrate the algorithm to correct SDSS DR12 data. 

\begin{figure}
    \centering
    \includegraphics[width=8cm]{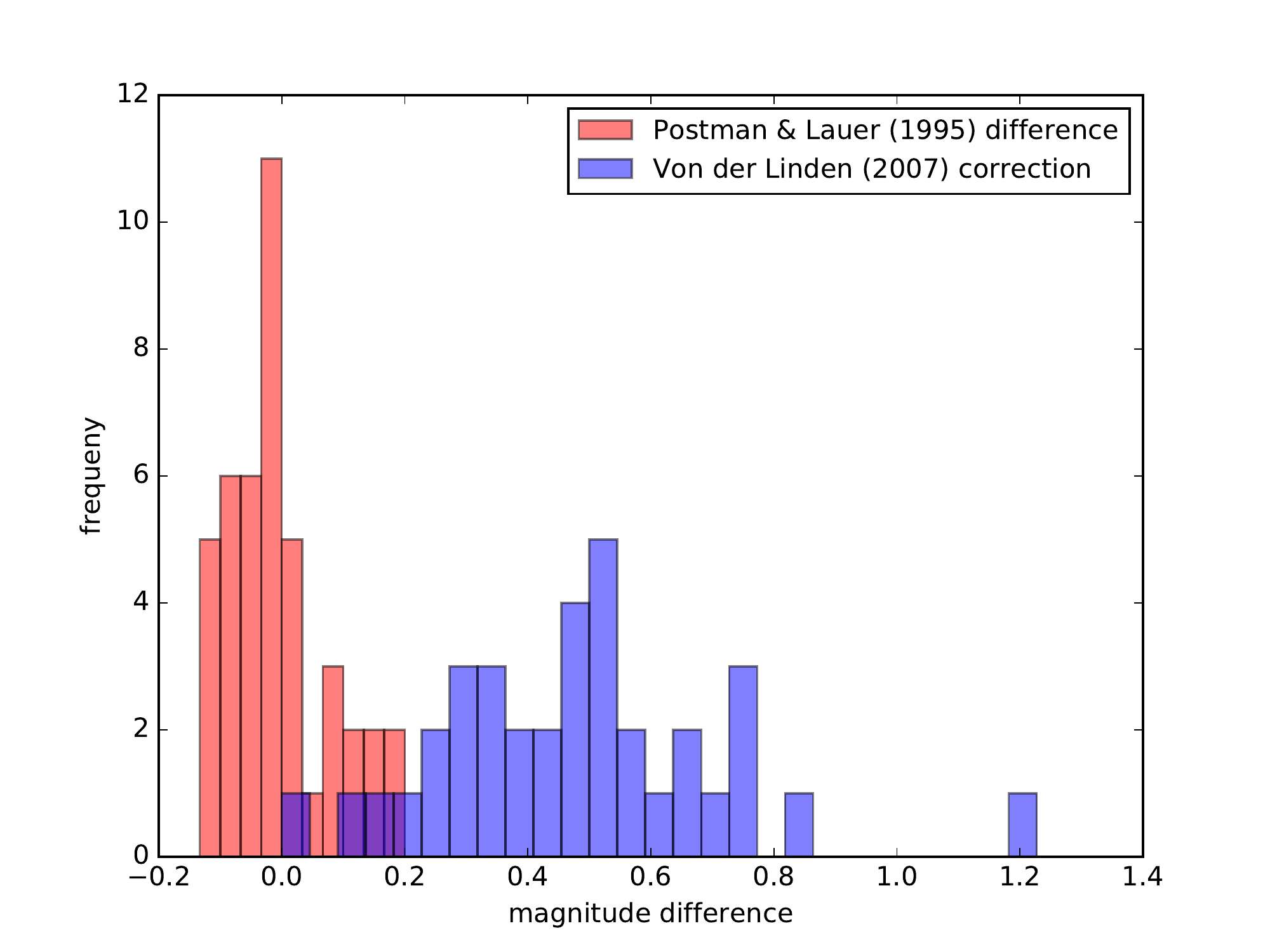}
    \caption{We compare our background corrected Petrosian BCG magnitudes to the total magnitudes in the \cite{pos95} sample, which is based on deeper imaging and has a more accurate background.  The red distribution shows the difference between our BCG Petrosian magnitudes and the \citet{pos95} magnitudes (within the Petrosian radius) after this correction is made.  We find that our algorithm can recover the \citet{pos95} magnitudes to within $\pm 0.1$ mags, which we define as the statistical floor of our BCG magnitudes.  The blue distribution shows the level of the correction to the SDSS Petrosian magnitude r-band that we apply based on our algorithm.  Of note, 81\% of the matches to the \citet{pos95} sample required a correction.}
    \label{fig:post_compare}
\end{figure}

\cite{pos95} utilize deep imaging to create precise light-profiles to measure BCG total magnitudes for a small subset of our cluster sample.  In Figure \ref{fig:post_compare}, we compare the results of our corrected BCG magnitudes against the \cite{pos95} BCG magnitudes. We find that our background corrected magnitudes recover the more carefully measured \cite{pos95} BCG magnitudes to within 0.1 magnitudes, which we define as the statistical floor for all of our BCG magnitudes.  Note that we only apply the re-calibrated \citet{von07} correction to our galaxies when the difference between the local sky background and the global sky background is positive and when the correction itself is greater than our baseline precision (0.1 mags) in each of the bands used in our final analysis (r and i).  

\begin{figure}
    \centering
    \includegraphics[width=8cm]{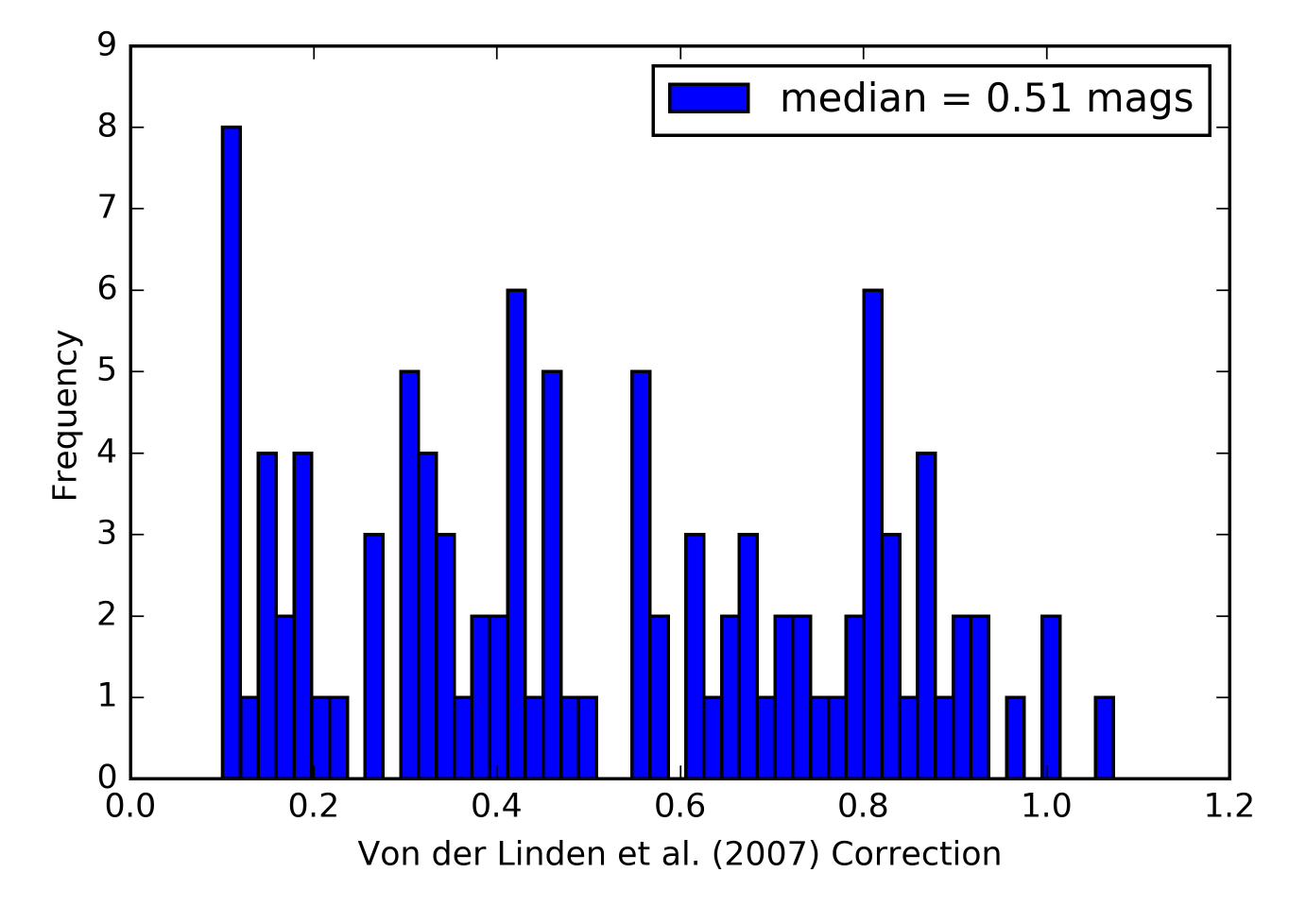}
    \caption{The distribution of the \citet{von07} r-band corrections $\ge$ 0.10 magnitudes for our SDSS-C4 BCGs.  The majority of corrections are large, which highlights the importance of background corrections prior to estimating stellar masses.}
    \label{fig:VonHist}
\end{figure}

Of our 370 BCGs, we find that 28\% require a \citet{von07} correction greater than our photometric accuracy, 0.1 magnitudes in both the r and i-bands.  Additionally, the median r-band \citet{von07} correction is 0.51 magnitudes as shown in Figure~\ref{fig:VonHist}, which underscores the importance of correcting the SDSS photometric measurements prior to estimating stellar masses.  Also, we find that the BCG magnitude corrections are uncorrelated with their magnitude gaps.

After applying the \citet{von07} correction to our SDSS light profiles, we used the cluster redshifts and apply a k-correction for the r and i-band magnitudes using the kcorrect (v 4.1.4) code \citep{bla07} to obtain our final BCG apparent magnitudes.  

To estimate stellar masses, we used the color-dependent M/L ratio from \citet{bel03} using the r-i color.\footnote{We only use the r-i color to estimate stellar mass, so the requirement that the \citet{von07} corrections be $\ge$ 0.1 magnitudes is only applied to the r and i-bands.  The g-band was not used in our final stellar mass estimates.}  Our choice of a color-dependent M/L ratio \citep{bel03} makes the stellar mass and associated error measurement used in our model, described in Section~\ref{sec:model}, dependent on the identification of ``red and dead'' elliptical galaxies.  To verify that this description matches our selected BCGs, we compare the color of each individual BCG to the color of a fiducial BCG, which we model using the EzGal SED modeling software \citep{man12}, assuming a \citet{bru03} SPS model, a \citet{cha03} IMF, a formation redshift of 4.9, and a redshift matching the individual BCG's.  However, neither the choice of IMF or z$_{form}$ strongly impacts the modeled color.  We confirm that our selected BCGs are ``red and dead'' if their measured color is within 0.1 magnitudes of the color of our fiducial BCG at the same redshift. 
 
We apply this criteria based on our fiducial BCG because the \citet{bel03} M/L ratio we use to estimate the stellar mass is color dependent.  Therefore, we remove BCGs whose colors do not match the fiducial BCG model colors because such colors resulted in non-physical stellar mass determinations based on our nominal SED analysis.  If we naively incorporate this data, we see strong deviations from Gaussianity, in that ~30 of the BCGs that don't match the fiducial model are outliers in the SMHM versus M14 relation.  Instead of selectively removing just the outliers, we chose to exclude all BCGs that do not match our fiducial model from the analysis, even though the majority fit the relation.  Another approach would be to fit an additional outlier component in our model and an even better solution would be to conduct a detailed SED modeling analysis of all of the BCGs, investigating which dust models and star formation histories best fit the colors.  Both of these approaches are beyond the scope of this effort.  Our sample is large enough to be selective against the quality of the SED model fits. We note that removing a random selection of 25\% of the BCGs of our final sample does not change our results.  Applying this color restriction removed 90 clusters, reducing our total sample to 280 clusters.  Therefore, roughly three-quarters of  the BCGs can be characterized as agreeing with our fiducial ``red and dead'' model; of the BCGs that were removed, 44\% are $ > 0.1$ magnitudes bluer and 56\% are $ > 0.1$ magnitudes redder than that of our fiducial BCG model. 

\subsection{Quantifying the Magnitude Gap}
\label{subsec:Maggap}
For our analysis, we chose to measure the magnitude gap between the BCG and fourth brightest cluster member (M14) and the BCG and second brightest cluster member (M12).  To do this, after identifying the BCG in each cluster, we identify red sequence galaxies within 0.5~$R_{vir}$ to determine the cluster membership and magnitude gap.  We used M14 and M12 as well as the 0.5~$R_{vir}$ for our membership and magnitude gap measurements because these were the magnitude gaps and radial extent used in the standard definitions of fossil group galaxies from \citet{jon03} and \citet{dar10}.  We fit the individual cluster red sequences in six distinct SDSS colors (u-g , g-r, g-i, i-r, i-z, and r-z) for all galaxies with an r-band magnitude brighter than $m_r$=19 within the 0.5 $R_{vir}$ region.

\begin{figure}
    \centering
    \includegraphics[width=8cm]{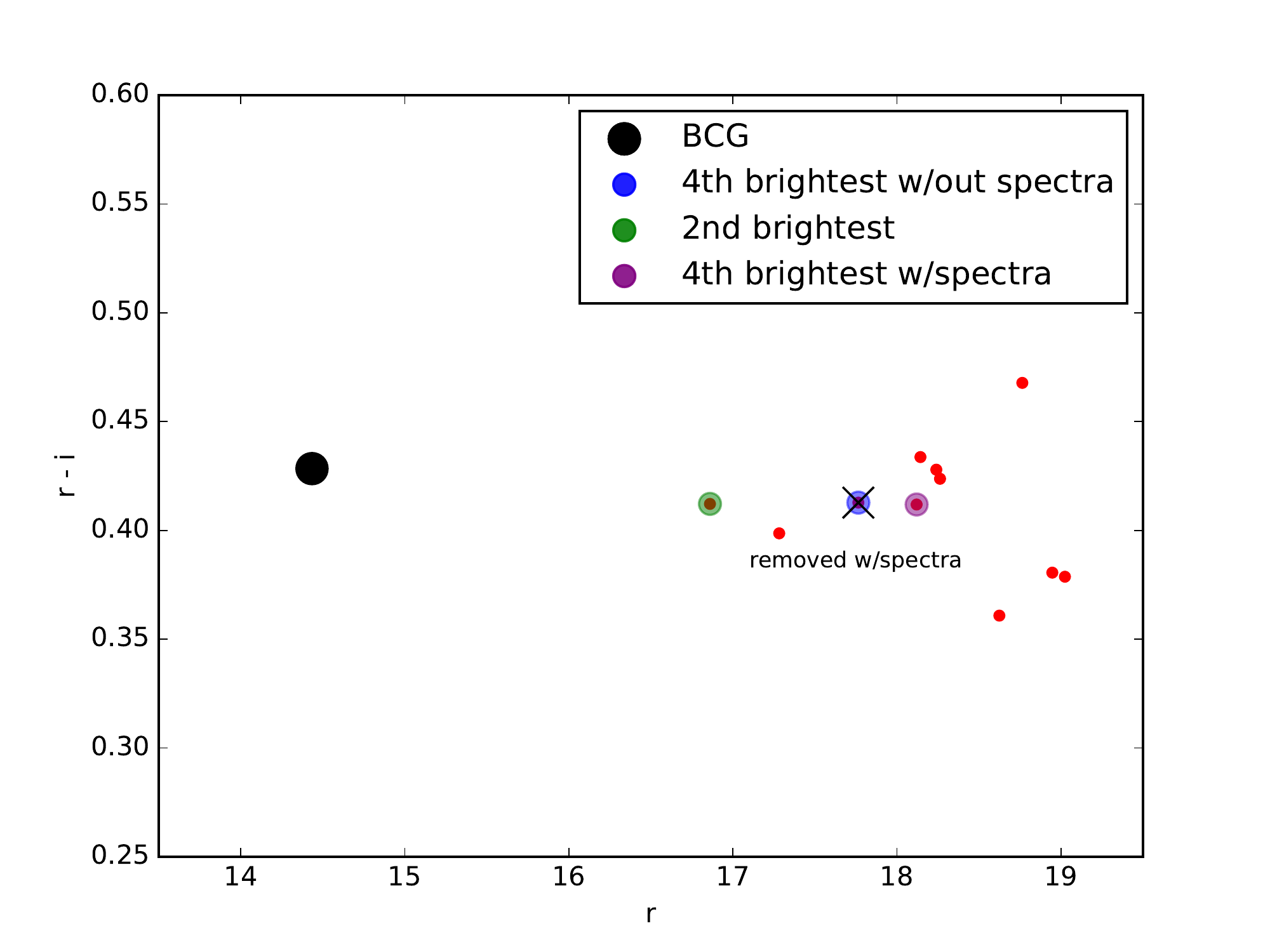}
    \caption{A sample color-magnitude diagram from one of our SDSS-C4 clusters.  This figure illustrates the impact of incorporating the spectroscopic redshift information.  The black point is the BCG, the green point is the 2nd brightest galaxy, the blue point is the 4th brightest galaxy when not using available spectroscopic information, and the purple point is the 4th brightest galaxy when using spectroscopic information.  This color magnitude diagram highlights that using the limited spectroscopic information further removes foreground contaminants, leading to a better identification of the 4th brightest galaxy in the cluster.}
    \label{fig:CMD-spec}
\end{figure}

We then characterize the magnitude gap using galaxies within $3\sigma$ of the fit to the red sequence for the u-g, g-r, and g-i colors and $2\sigma$ for the i-r, i-z, r-z colors.  We note that we do not require galaxy spectroscopy for cluster membership; however, whenever possible, we do utilize available SDSS spectroscopic redshifts and remove any potential cluster member with $|z_{gal} - z_{clus}| > 3 \sigma/{c}$.  In Figure~\ref{fig:CMD-spec}, we present an example color magnitude diagram for one of our clusters to highlight the impact of incorporating the available spectroscopic redshifts to further remove remaining foreground contaminants (the point with an x through it in Figure~\ref{fig:CMD-spec}).  Doing so leads to a fainter 4th brightest galaxy and an increase in the magnitude gap. 
We note that spectroscopic information is not available for all red sequence candidates and that the incorporation of available spectroscopic redshifts only changes the 4th brightest galaxy for 5\% of our clusters.  We further address the use of galaxy redshifts and red-sequence membership in Section~\ref{sec:Sims}, where we use a simulated sky survey to constrain systematic uncertainties due to projection effects in the magnitude gaps.  Additionally, we do not require the BCG to have a spectrum to place it within the spectroscopically confirmed cluster.  Since the measurement of the magnitude gap is dependent upon fitting a red sequence, as previously noted, we remove the 12 clusters from our sample which either had fewer than 4 members or for which we are unable to fit a red sequence for the cluster members within 0.5 $R_{vir}$.  

After we identified the second and fourth brightest cluster members, we apply k-corrections using the cluster redshifts and kcorrect (v 4.1.4) code \citep{bla07} to obtain our final fourth and second brightest member apparent magnitudes.  We then measure the two unique magnitude gaps, M14 and M12, in the r-band.  Unlike the BCGs the majority of second and fourth brightest galaxies are neither extended nor located in dense regions; therefore, we do not apply a \citet{von07} correction.

Using the final sample of 280 clusters, we ran a completeness analysis following a similar approach to what is described in \citet{col89}, \citet{gar99}, \citet{lab10}, and \citet{tre16}.  To determine the completeness, we first convert the apparent magnitudes for our BCGs and 4th (2nd) brightest members to R-band absolute magnitudes, then bin the absolute magnitudes by apparent magnitude, and calculate the 95\% limit in each bin.  Using this upper limit, we fit a linear relation between the upper limit and apparent magnitude of each bin and determine the absolute magnitude that corresponds to an apparent magnitude of $m_r=19$, the upper limit we applied to our SDSS galaxy catalogs.  We chose this rather bright upper limit to minimize additional photometric measurement uncertainties in our analysis.  We note that the absolute magnitude limit for the 2nd brightest galaxies is dimmer than the limit for the 4th brightest because we require at least 4 members in the red-sequence.    

\begin{figure}
    \centering
    \includegraphics[width=8cm]{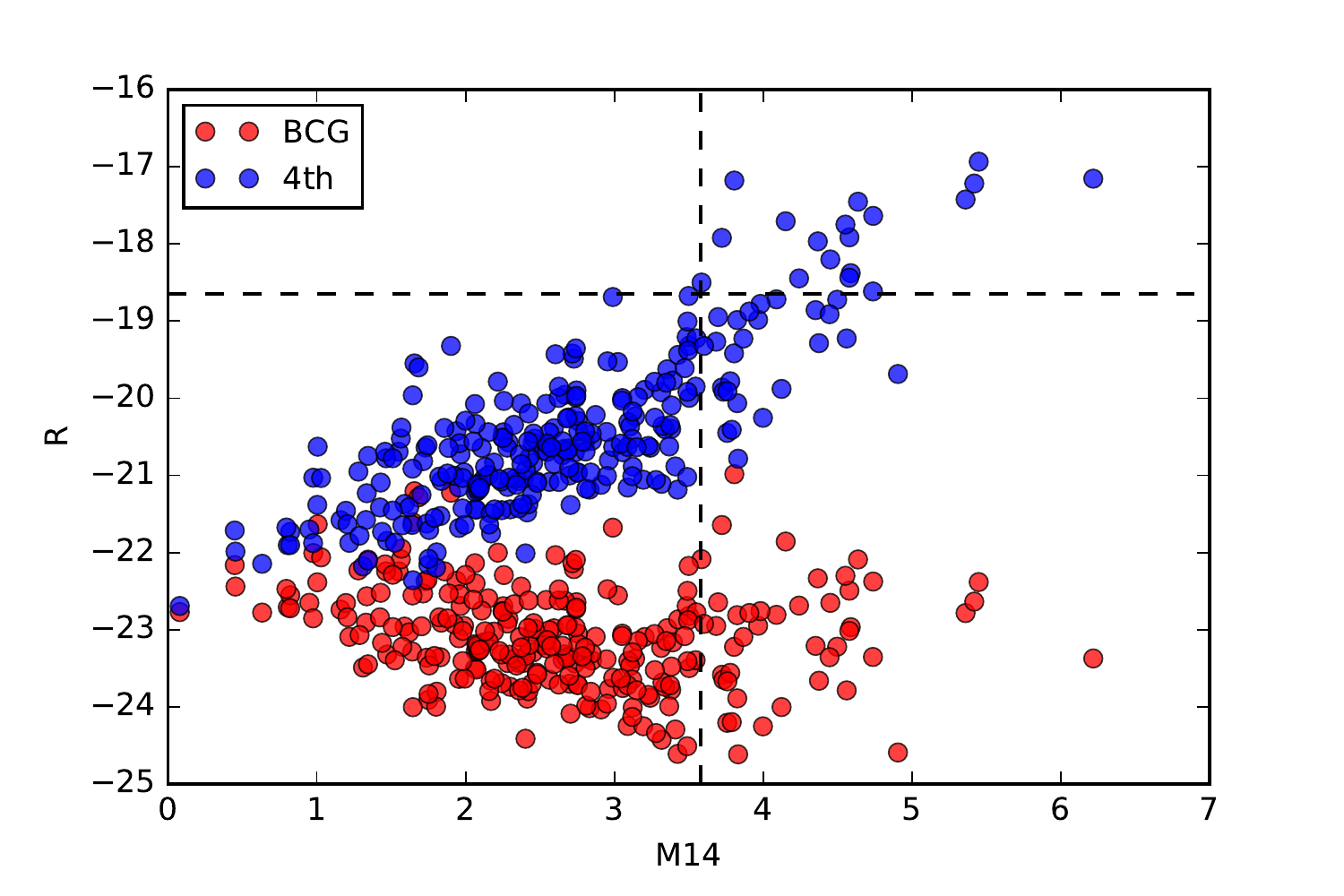}
    \caption{R-band absolute magnitude versus M14. This is used to determine the completeness of our SDSS-C4 sample.  The vertical and horizontal lines represent the limits in M14 and absolute magnitude, respectively.  The BCGs are shown in red and the 4th brightest cluster members are shown in blue.}
    \label{fig:Comp-M14}
\end{figure}

\begin{figure}
    \centering
    \includegraphics[width=8cm]{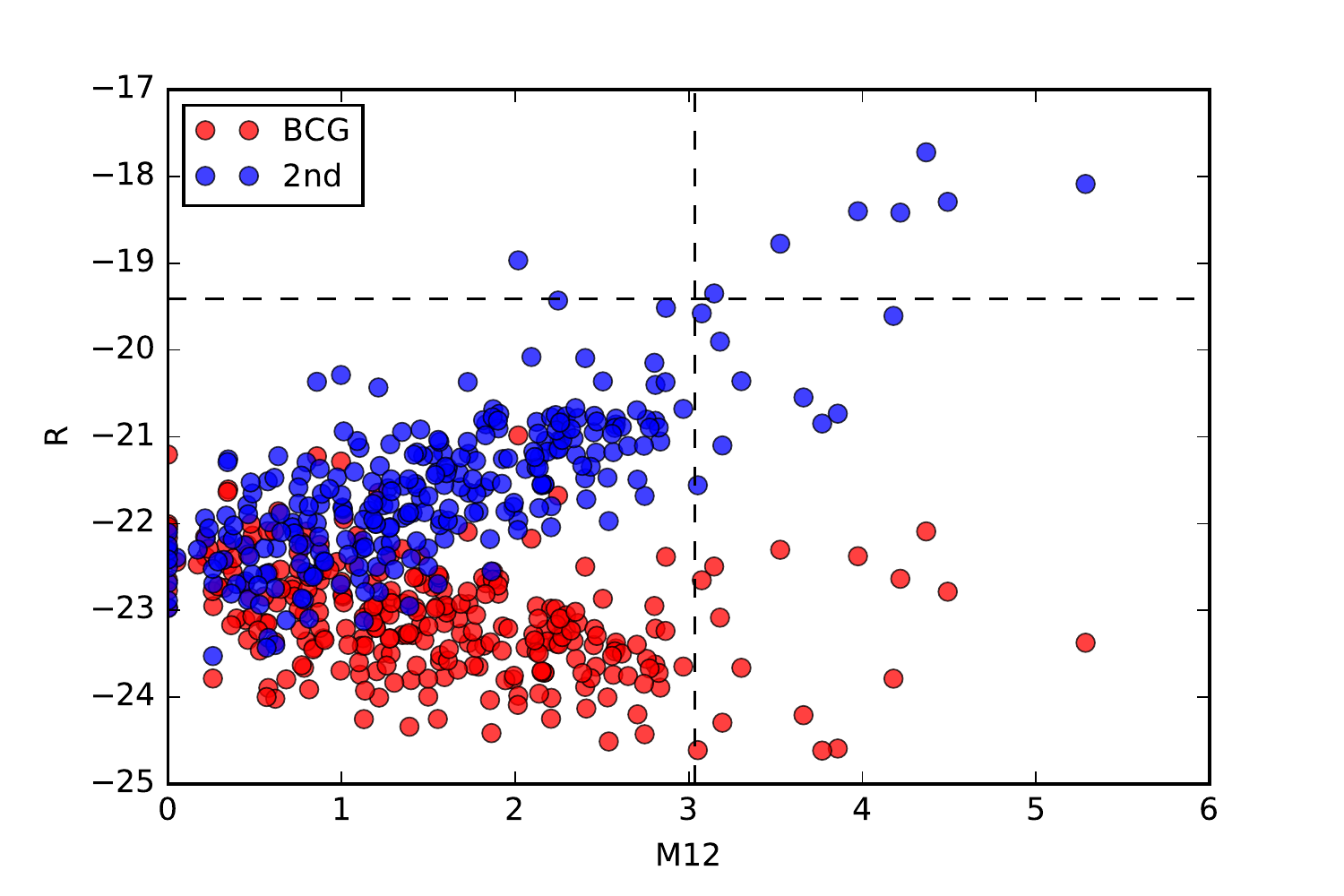}
    \caption{R-band absolute magnitude plotted against M12.  This is used to determine the completeness of our SDSS-C4 sample.  The vertical and horizontal lines represent the upper limits in M12 and absolute magnitude, respectively.  The BCGs are shown in red and the 2nd brightest cluster members are shown in blue.}
    \label{fig:Comp-M12}
\end{figure}

Next, we follow the same procedure, but instead bin the absolute magnitude as a function of magnitude gap and determine the 95\% limit in each bin, which leads to an upper limit on the M14 (M12) magnitude gap of 3.58 (3.035).  After applying these cuts, we remove 44 (26) additional clusters, as shown in Figures~\ref{fig:Comp-M14} and \ref{fig:Comp-M12}, leaving us with a total sample of 236 (254) ``red and dead'' elliptical BCGs with magnitude gap measurements.  If we apply more stringent cutoffs for both the magnitude gap and absolute magnitude, we find no significant difference in the posterior distributions. 

\subsection{Final Sample Summary}
\label{subsection:Final_Sample}
After applying the different criteria, summarized in Table~\ref{tab:SDSS-C4-data} we are left with our final sample which contains either 236 or 254 clusters, depending on the magnitude gap used.  Our sample can be characterized as having $log10(M_{halo}/h) \ge 14.0$, which are measured using the caustic technique on well sampled radius-velocity phase spaces.  The halo mass uncertainties of the clusters in our final sample range from 0.31 - 0.46 dex with a median of 0.32 dex.  For each cluster the BCG is a ``red and dead'' elliptical galaxy with a color that is within 0.1 magnitudes of the color of our fiducial BCG.  Moreover, each cluster has a clearly defined red sequences with greater than four members, including 2nd or 4th brightest galaxies whose absolute magnitudes and associated magnitude gaps fall within our completeness criteria.

\begin{deluxetable*}{ccc}
	\tablecaption{SDSS-C4 Final Sample: Summary of removed clusters}
	\tablecolumns{3}
	\tablewidth{0pt}
	\tablehead{\colhead{Selection Criteria} & 
	\colhead{Number Removed} & 
	\colhead{Number Remaining}
	} 
\startdata
C4 clusters & & 970 \\
$log10(M_{caustic}/h) \ge 14.0$ & 450 & 420 \\
Broad unpeaked velocity histogram, unidentifiable red sequence, poorly defined caustic & 32 & 388 \\
Photometric errors & 6 & 382 \\
No red sequence within 0.5$R_{vir}$ & 12 & 370 \\
BCG color does not match ''red and dead'' fiducial BCG & 90 & 280 \\
Completeness analysis & 44 (26) & 236 (254) \footnote{This represents the results of the completeness analysis for the sample done using the M14 (M12) magnitude gaps.} \\
Final sample & & 236 (254) \\

\enddata
\label{tab:SDSS-C4-data}
\end{deluxetable*}

\section{Simulated Data}
\label{sec:Sims}
In addition to studying the impact of the magnitude gap on the SMHM relation in the SDSS-C4 cluster sample, we also analyzed this trend in simulations using the \citet{guo10} and \citet{hen12} prescriptions of the semi-analytic representations of low-redshift galaxy clusters in the MILLENNIUM simulation.  We chose these semi-analytic models because the galaxies grow hierarchically, which as \citet{solanes16} suggest, may relate to the magnitude gap.  We note that the \citet{hen12} prescription is constructed by taking the semi-analytic redshift snapshots from \citet{guo10} and stitching them together to create a mock light-cone \citep{hen12}.  We use both the 3D information from this light-cone as well as the full projected data in our analyses that follow.   

In our analysis of the simulations, we aim to treat the simulated data in a similar manner to the SDSS-C4 observational measurements of stellar mass, halo mass, and magnitude gap.  In doing so we generate two catalogs; a sample that uses all of the 3D information provided for the cluster (i.e., halo masses measured within $r_{200}\times$ $\rho_{crit}$, galaxy positions in x, y, z, and semi-analytic stellar masses, and magnitudes) and a projected sample, which instead uses only 2D information (i.e., caustic-inferred cluster masses, RA, Dec, redshift, apparent magnitudes, and inferred stellar masses).  

When using the 3D data, we use the stellar masses for each central halo's galaxy directly from \cite{hen12}.  For both the 3D and 2D data, we identify the BCG as the brightest red-sequence galaxy within 0.5$R_{vir}$. We then use the red-sequence to define the magnitude gaps.  We estimate the BCG stellar masses for the 2D sample using the same \citet{bel03} M/L ratio conversion based on the r-i color.  Finally, the projected cluster masses were determined using the caustic technique (2D) \citep{gif13a}.  In other words, the 2D projected data are treated almost identically to the real SDSS data, except for visual classification and identification of the BCG.

Determining cluster membership and the magnitude gap was done slightly differently for our simulations because we have different available information.  For the 3D sample, we use the positional information (x, y, z) to determine if a galaxy is within 0.5~$R_{sim,vir}$.  For the galaxies within this sphere, we use the red sequence to determine cluster membership.  For the 2D projected sample, we followed the same steps outlined for our own observations.  We use RA and Dec, to determine if a galaxy is within a circle of 0.5~$R_{vir}$ centered on the BCG.  Next we check to see if $|z_{gal} - z_{clus}| < 3 \sigma/{c}$, where $\sigma$ is the measured velocity dispersion.  For those remaining cluster members, we use the red sequence to determine membership.  We note that the measurements in our 2D projected catalog introduce error in the halo mass, from the caustic measurements, stellar mass, and magnitude gap.  Understanding how to incorporate these additional errors was crucial in creating our Bayesian MCMC model for the SMHM relation, as described in Section~\ref{sec:model}. 

Of note, unlike with our observations, the simulated BCGs are treated as being ``red and dead''.  Furthermore, because we have access to the complete 3D simulation box, we do not perform a completeness test to remove some of the clusters with either extremely large magnitude gaps or fainter second and fourth brightest galaxies.  Additionally, we apply the same mass thresholds on both the 2D and 3D simulations and the SDSS-C4 data such that the observed mass is $log(M_{caustic}/h) \ge 14.0$ and the underlying halo mass distribution for the simulated data is an approximation of the \citet{hen12} mass function truncated at $log(M_{halo}/h) \ge 14.0$.  We note that if we adjust this halo mass thresholds by 0.1 dex in either direction the results of the posterior distributions are within one sigma of those presented in Figures~\ref{fig:SMHMR-HenT} and \ref{fig:SMHMR-HenP}.  

\section{The Hierarchical Bayesian Model}
\label{sec:model}
Unlike other quantitative analyses of the SMHM relation \citep[e.g.,][]{yan09, mos10, mos13}, we do not use a two component power law to describe our SMHM relation.  Instead, we use a single component power law, because we are only concerned with the high mass portion, $log(M_{halo}/h) \ge 14.0$, of the SMHM relation.  Additionally, we do not allow for redshift evolution in this low-redshift dataset ($z \le 0.18$, $z_{median} = 0.086$) unlike other published SMHM relations \citep[e.g.,][]{beh13,mos13} because the results of these prior studies suggest the change of the slope of the SMHM relation over this redshift range is smaller than the precision of the posteriors for the slope of our Bayesian MCMC analysis.

We assume a linear model for our data, given by Equation~\ref{eq:linear_model},
\begin{equation}
\label{eq:linear_model}
    log10(M_{*})=\alpha + \beta (log10(M_{halo})) + \gamma (M14)
\end{equation}
which is parameterized by a slope, $\beta$, y-intercept, $\alpha$, and the ``stretch'' parameter related to the magnitude gap, $\gamma$.  For the remainder of the analysis, we refer to the log (base 10) BCG stellar masses as $y$, the log (base 10) cluster (or halo) masses as $x$, and the magnitude gap as $z$. 

To determine the appropriate values for $\alpha$, $\beta$, and $\gamma$, as well as the intrinsic scatter, $\sigma_{int}$, in our relation, we use a Bayesian MCMC analysis to maximize the sum of the log-likelihoods for each cluster.  The Bayesian approach can briefly be described as convolving the prior information for a given model with the likelihood of the observations given the model, which yields the posterior distribution, or the probability of observing the data given the model.  We use a Bayesian analysis because it allows us to easily account for all prior information.  It is hierachical in the sense that we model both the errors on our parameters, as well as the uncertainty on those errors.

Throughout our Bayesian likelihood analysis, the MCMC model generates values for stellar mass, halo mass, and magnitude gap, which are directly compared to the respective observed measurements given the errors in the data.  The comparison between the model generated data and our observations allows us to construct posterior distributions for all of our free parameters, from which we can determine the most likely values for $\alpha$, $\beta$, $\gamma$, and $\sigma_{int}$ as shown in the triangle plots presented in Sections~\ref{subsubsec:MILL-QI} and \ref{subsubsec:R-QI}.  In the following subsections and Table \ref{tab:bayes}, we provide all of the details regarding our model.

\subsection{The Observed Quantities}
\label{subsec:ObservedQuantities}
We use the observed BCG magnitudes and colors to create a stellar mass, $y_{0i}$, which we treat as an observable. There is also an error on the observed stellar masses, which we treat as Gaussian, so the observed stellar mass is:
\begin{equation}
y_{0i} \sim \mathcal{N}(y_i,\,\sigma_{y_i}^{2}).
\end{equation}
where $y_i$ is the underlying (and unknown) true stellar mass and $\sigma_{y_i}$ is the modeled uncertainty on the measurement.  We model the observed uncertainty on the stellar mass as a beta distribution centered on a measured value, $\sigma_{y_{0i}}$.  This value can vary depending on whether we are analyzing the real or the simulated data (more details below). The use of the beta distribution allows for stochasticity and additional uncertainty (e.g., systematic bias) in our observed stellar mass errors.
\begin{equation}
\sigma_{y_{i}} = \sigma_{y_{0i}} \pm \mathcal{B}(a,b)
\label{eq:beta_y}
\end{equation}
where $a=0.5$ and $b=100$ are the shape parameters of the beta distribution. These shape parameters add up to $\pm 0.06$ dex to our BCG stellar mass error estimates. 

For the observed cluster masses, we use the observed radius-velocity phase-spaces to produce observed caustic masses. We apply a lower mass threshold of $1\times10^{14}$M$_{\odot}/h$ and thus use a truncated normal defining the cluster caustic masses:
\begin{equation}
x_{0i} \sim {\rm T}\mathcal{N}(x_{i},\,\sigma_{x_{i}}^{2}, x_{0min}, x_{0max}).
\end{equation}
where $x_{i}$ is the underlying (and unknown) true halo mass, $\sigma_{x_{i}}$ is the modeled uncertainty on the mass measurement and $x_{0min}$ and $x_{0max}$ specify the lower and upper limits on the observed caustic masses in the sample.  We note that for the simulations, we apply the same observational halo mass limit.  In practice, for the SDSS-C4 data, we use the ``observed'' uncertainty, $\sigma_{x_{0i}}$, on each cluster mass, which is based on the mapping between the number of galaxies in the phase-space to the caustic mass uncertainty as quantified in simulations \citep{gif13a}.  Because there is some uncertainty due to the use of simulations, we add an additional stochastic component drawn from the beta probability distribution:
\begin{equation}
\sigma_{x_{i}} = \sigma_{x_{0i}}(N_{phase_i}) \pm \mathcal{B}(a,b)
\label{eq:beta_x}
\end{equation}
where $a=0.5$ and $b=100$ are the shape parameters of the beta distribution.  As before, this additional uncertainty ranges up to $\pm 0.06$ dex.

We treat the observed magnitude gap similarly to the halo masses.  The observed gap is drawn from a truncated Normal distribution:
\begin{equation}
z_{0i} \sim T\mathcal{N}(z_i,\sigma_{z_i}^2,z_{0min}, z_{0max}).
\end{equation}
where $z_i$ is the underlying (and unknown) true magnitude gap, $\sigma_z$ is the modeled uncertainty on the magnitude gap, and $z_{0min}$ and $z_{0max}$ represent the lower and upper limits on the magnitude gap, set by the completeness analysis.  We note that for the simulated data, we do not use a truncated Gaussian because we do not apply the completeness limits, as described in Section~\ref{sec:Sims}.  For the uncertainties on the magnitude gap, we compared the distribution of our ``observed'' gaps to the distribution of true 3D gaps using the simulations.  Additionally, for our SDSS-C4 clusters, we add an uncertainty, $\sigma_{z_{0i}}$, of 0.1 mags to account for the photometric measurement uncertainty of our BCG magnitudes.  Since this is also calibrated using simulations, we include an additional stochastic component using the beta probability distribution.
\begin{equation}
\sigma_{z_{i}} = \sigma_{z_0i} \pm \mathcal{B}(a,b)
\label{eq:beta_z}
\end{equation}
where $a=0.5$ and $b=100$ are the shape parameters of the beta distribution.  Like before, this addition ranges up to $\pm 0.06$ to the error in magnitude gap.  

Given the above likelihoods for an observed BCG ($x_{0i}$, $y_{0i}$, and $z_{0i}$), we sum the log of the likelihoods defined by the $\chi^2$ distribution since all of our probability distributions are Gaussian.  We will map the posterior using an MCMC approach.  However, we still need to define the unobserved parameters and their prior probability distributions.

\subsection{The Unobserved Quantities}

For each cluster $i$, our model returns a true log stellar mass ($y_i$) given the true magnitude gap ($z_i$), the true log halo mass ($x_i$), and the parameters that relate them ($\alpha, \beta, \gamma$).  We also allow for intrinsic scatter, $\sigma_{int}$, in our relationship. Thus, we model $y_i$ as:
\begin{equation}
y_i = \mathcal{N}(\alpha + \beta x_{i} + \gamma z_{i}, \sigma_{int}^2).
\label{eq:smhm_relation}
\end{equation}
We note in Equation~\ref{eq:smhm_relation} that the halo masses, the magnitude gaps, and the parameters $\alpha$, $\beta$, and $\gamma$ which together yield a BCG stellar mass are the true values.  The Bayesian model regresses against the observed stellar masses, cluster masses, and magnitude gaps self-consistently.  We define the relationship as being {\it causal} and relating the true underlying BCG stellar mass to the halo mass and the magnitude gap.  The parameters that we are interested in are the intercept, slope, stretch, and intrinsic scatter.  All other parameters are treated as nuisance parameters and are marginalized over when we present the posterior probability distributions.

We do not use uniform priors on the halo masses or the magnitude gaps because there is no reason to expect it in the real universe.  The prior on halo masses $x_i$, is the mass function of the halos given from the 3D \citet{hen12} data, modeled as a truncated Gaussian with a mean and width, which are treated as free parameters that are given by a fit to this mass function.
For this analysis, we truncate the halo mass function at $x_{i} = 14.0$. We find no significant difference in our posteriors if we lower this threshold to $x_{i} = 13.8$. In other words, our results are not sensitive to the exact truncation limit we use on the underlying mass function. 
The prior on the magnitude gaps is defined by the observed magnitude gap distribution modeled as a Gaussian where the mean and width are free parameters given by this magnitude gap distribution.  For the means and widths of both the magnitude gap and halo mass, the free parameters are modeled as Normal distributions. \footnote{For simplicity, in Table~\ref{tab:bayes} the value for the mean and width of $x_{i}$ and $z_{i}$ are our initial values for these modeled parameters.  The values given by the Normal distributions at each step, which can differ from the initial values, are used in our Bayesian MCMC analysis.  We note that our results are not sensitive to details of the initial values for either the halo masses or magnitude gaps.} 

In terms of the measurement uncertainties, our priors change depending on whether we are analyzing the simulations (3D or 2D) or the real data. For the 3D simulations, we assume there is zero uncertainty in the stellar masses, the halo masses, and the magnitude gaps.  For the 2D simulated data, we use $\sigma_y = 0.03$, which we determine by measuring the scatter in the difference between true stellar masses and the inferred stellar masses from the \citet{bel03} relation.  The 2D phase-spaces are better sampled than the observed data, and so we use a uniform value close to the lower limit (0.35 dex) of the scatter in the caustic mass from \cite{gif13a}.  The error in the magnitude gap for the simulated data is again negligible, since the dominant component (the photometric error) is not present.

We use uninformative uniform priors on the intrinsic scatter, $\sigma_{int}$, and on the intercept, $\alpha$.  For the slope, $\beta$, and the stretch factor, $\gamma$, we use a linear regression prior of the form $-1.5\times \rm{log}(1+value^2)$. 

In Table \ref{tab:bayes}, we summarize all of our likelihood forms and priors not previously described in Section~\ref{subsec:ObservedQuantities}.

\begin{widetext}
We can express the entire posterior probability then as:
\begin{equation}
\begin{aligned}
p(\alpha,\beta,\gamma,\sigma_{int}, x_{i},z_{i},\sigma_{y_i},\sigma_{x_i}, \sigma_{z_i}|y_{0i},x_{0i},z_{0i}, \sigma_{y_{0i}},\sigma_{x_{0i}},\sigma_{z_{0i}}) \propto & \\
& \underbrace{P(y_{0i}|\alpha,\beta,\gamma,\sigma_{y_i}, \sigma_{int}, x_{i},z_{i}) ~ P(x_{0i}|x_{i},\sigma_{x_i}) ~ P(z_{0i}|z_{i},\sigma_{z_i})}_{\text{likelihood}} \times \\ 
&  \underbrace{p(x_i) ~ p(z_{i}) ~  p(\sigma_{x_i}) ~ p(\sigma_{y_i}) ~ p(\sigma_{z_i}) ~
p(\alpha) ~ p(\beta) ~ p(\gamma) ~ p(\sigma_{int})}_{\text{priors}} 
\end{aligned}
\label{eq:posterior}
\end{equation}
\end{widetext}
We note that this is actually a {\it hierarchical Bayes model}, since the priors on true halo masses and true magnitude gaps ($x_i$ and $z_i$) depend on models themselves (e.g., the underlying halo mass function or a distribution of the magnitude gap data).  We do not include those terms in equation \ref{eq:posterior} or in Table \ref{tab:bayes} for compactness, but they are described in the above text.  Additionally, since we allow for errors in all of our observables, this is an extension of what was originally shown for a simpler Bayesian line-fitting analysis in \citet{gull89}.

\begin{deluxetable*}{ccc}
	\tablecaption{Bayesian Analysis Parameters for the SDSS-C4 Nominal Sample using M14}
	\tablecolumns{3}
	\tablewidth{0pt}
	\tablehead{\colhead{Symbol} & 
	\colhead{Description} & 
	\colhead{Prior}
	} 
\startdata
$\alpha$ & The offset of the SMHM relation & $\mathcal{U}$(-100,100) \\
$\beta$ & The high-mass power law slope & Linear Regression Prior \\
$\gamma$ & The stretch factor, which relates to M14 & Linear Regression Prior \\
$\sigma_{int}$ & The uncertainty given by the width of the 
intrinsic stellar mass distribution & $\mathcal{U}(0.0,0.5)$\\
$y_{i}$ & The underlying distribution in stellar mass given by Equation~\ref{eq:smhm_relation} & $\mathcal{N}$($\alpha + \beta x_{i} + \gamma z_{i} $,$\sigma_{int}^2$)\\
$x_{i}$ & The underlying halo mass function from \citet{hen12}, approximated as a truncated Normal & T$\mathcal{N}$(14.0,$0.35^2$,14.0,15.1)\\
$z_{i}$ & The underlying magnitude gap distribution & $\mathcal{N}$(2.4,$0.71^2$)\\
$\sigma_{y_{0i}}$ & The uncertainty between the observed stellar mass and intrinsic stellar mass distribution & 0.19\\
$\sigma_{x_{0i}}$ & The uncertainty between the caustic halo mass and underlying distribution given by \citet{gif13a} & $\sigma_{x_{0i}}(N_{phase_i})$\\
$\sigma_{z_{0i}}$ & The uncertainty between the underlying and observed halo mass distribution & 0.10\\
  & & \\
\caption{$\mathcal{U}(a,b)$ refers to a uniform distribution where a and b are the upper and lower limits.  The linear regression prior is of the form $-1.5 \times log(1+value^2)$.  $\mathcal{N}(a,b)$ refers to a Normal distribution with mean and precision of a and b.  T$\mathcal{N}(a,b,c,d)$ is a truncated Normal distribution with mean, a, precision, b, lower limit, c, and upper limit, d.  For $\sigma_{x_{i}}$, the value is dependent on the number of galaxies used in the caustic phase space and is obtained from \citet{gif13a}.  Additionally, we note that for $x_{i}$ and $z_{i}$, the means and widths given in this table are the initial values.  The mean and widths are modeled as normal distributions whose values are used at each step in the Bayesian MCMC analysis.  The results are not sensitive to those initial values.} 
\enddata
\label{tab:bayes}
\end{deluxetable*}

\section{Results}\label{sec:Results}
In this section, we present the SMHM relation incorporating either M14 or M12 and the results of our Bayesian MCMC model for both the simulated and the SDSS-C4 observed data.

\subsection{MILLENNIUM Simulation} \label{subsec:R-MILL}
\subsubsection{SMHM Relation}\label{subsubsec:MILL-SMHM}
Here we present the qualitative results of our analysis of the SMHM relation for high mass clusters found in the 3D and 2D (projected) versions of the \citet{hen12} prescription of the MILLENNIUM simulation. 

\begin{figure}
    \centering
    \includegraphics[width=8cm]{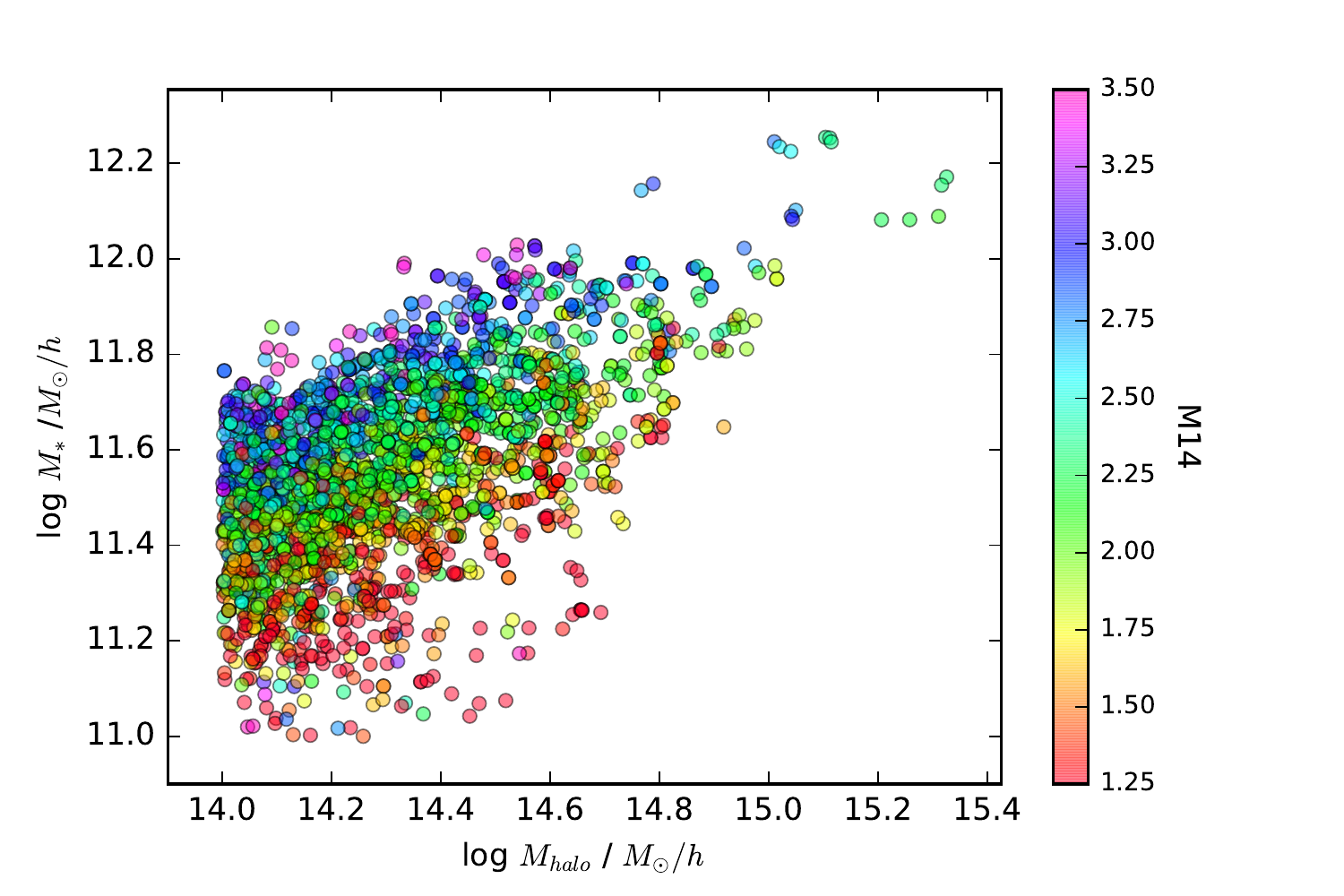}
    \caption{The SMHM relation for the 3D sample of the \citet{hen12} prescription of the MILLENNIUM simulation.  In the simulated universe we see a smooth magnitude gap -- stellar mass stratification.}
    \label{fig:SMHMR-HenT}
\end{figure}
Figure~\ref{fig:SMHMR-HenT} shows the stellar masses plotted against the halo masses, which both come directly from the simulation, for the 3D cluster sample.  The BCGs are color-coded, where red represents small magnitude gap BCGs and purple high-gap BCGs (for M14).
\begin{figure}
    \centering
    \includegraphics[width=8cm]{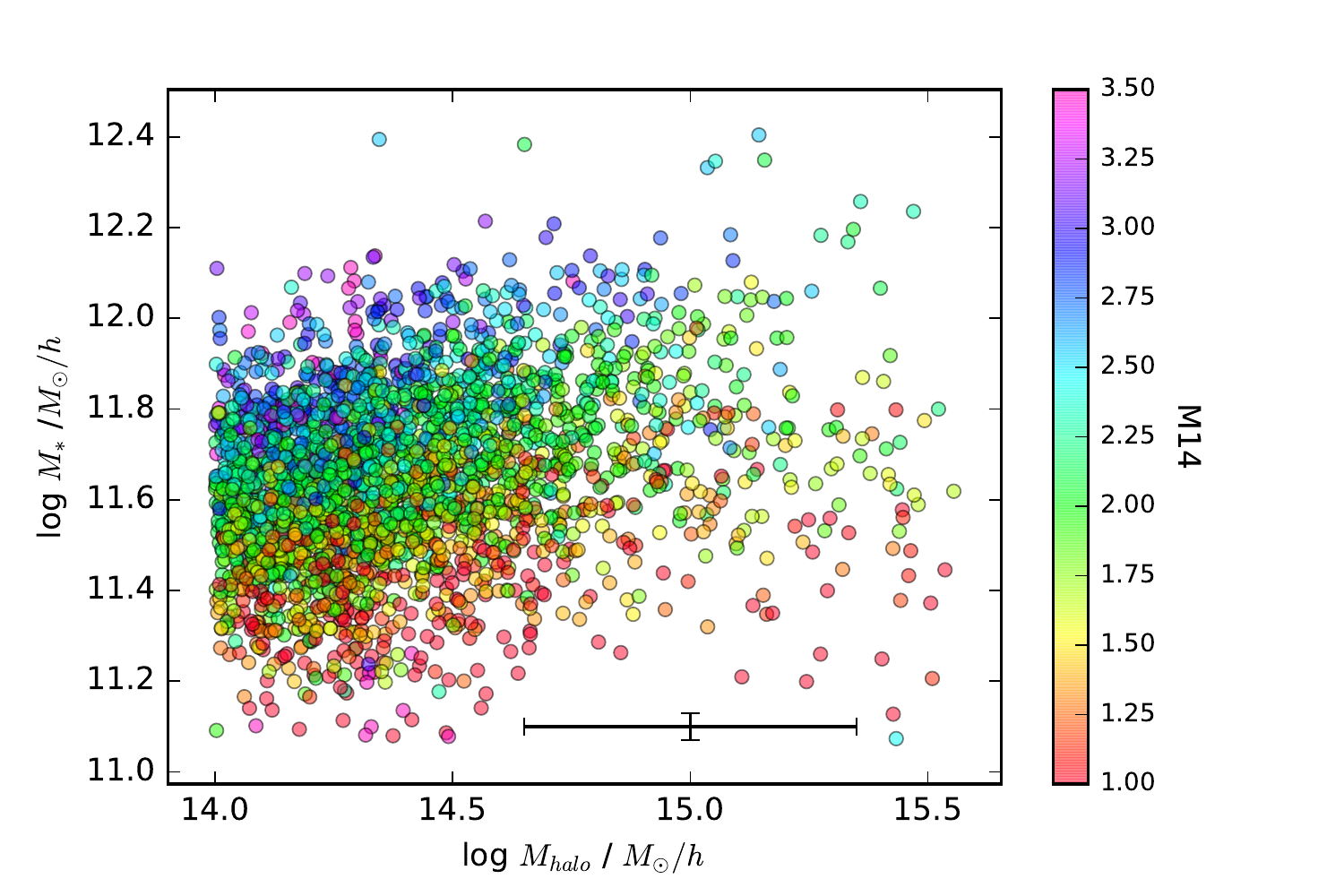}
    \caption{The SMHM relation for the 2D projected sample of the \citet{hen12} prescription of the MILLENNIUM simulation.  The black bar shows the average error in halo mass, 0.35 dex, and stellar mass, 0.03 dex.  Even when using data with measurement errors in stellar mass, halo mass, and magnitude gap, a similarly smooth magnitude gap -- stellar mass stratification exists in the simulated data, like in Figure~\ref{fig:SMHMR-HenT}.}
    \label{fig:SMHMR-HenP}
\end{figure}
Figure~\ref{fig:SMHMR-HenP} presents the SMHM relation for the 2D projected sample, where the stellar masses are estimated using the \citet{bel03} M/L ratio relation, and the halo masses are determined using the caustic technique on the projected phase-spaces.  Again, the BCGs are color-coded according to their magnitude gap (M14), which is determined using the projected data.

For Figures~\ref{fig:SMHMR-HenT} and \ref{fig:SMHMR-HenP}, the magnitude gap color bar does not span the entire range of observed M14 values (0-4.3).  Instead, since the M14 distribution can be approximated as a Gaussian centered at an M14 value of $\sim$2.1, we selected a range which eliminates the Gaussian wings to better highlight the difference in stellar mass at fixed halo mass for clusters with differing magnitude gaps. 

Figures~\ref{fig:SMHMR-HenT} and~\ref{fig:SMHMR-HenP} show that a relationship clearly exists between the magnitude gap and stellar mass at a fixed halo mass in the semi-analytic prescription of the MILLENNIUM simulation.  Recall that \citet{har12} identified a bifurcation between clusters with high magnitude gaps and low magnitude gaps; high gap clusters have a larger stellar mass at fixed halo mass than low gap clusters.  Our analysis illustrates that in the MILLENNIUM simulation, there is a continuous stratification in the BCG stellar masses at fixed halo mass due to the magnitude gap.  These qualitative results are unchanged when using M12 (not shown) instead of M14.
 
\subsubsection{Quantitative Impact} \label{subsubsec:MILL-QI}
To quantitatively evaluate if the magnitude gap can be treated as a latent parameter in the SMHM relation, we use our MCMC model, Bayesian formalism, and linear SMHM relation (Equation~\ref{eq:smhm_relation}) described in Section~\ref{sec:model}.  To convey these results, we present triangle plots that show the posterior distributions of $\alpha$, $\beta$, $\gamma$, and $\sigma_{int}$ plotted against one another for both the 3D and 2D \citet{hen12} samples.  Each of these plots is generated after 10 million steps (including an approximate 2 million step burn in).  We marginalize over all other nuisance parameters from Equation~\ref{eq:posterior}.  We present the results from the \citet{hen12} simulations using M14 for the magnitude gap.  All of the results from the posterior distributions are presented in Table~\ref{tab:results}.

\begin{figure}
    \centering
    \includegraphics[width=8cm]{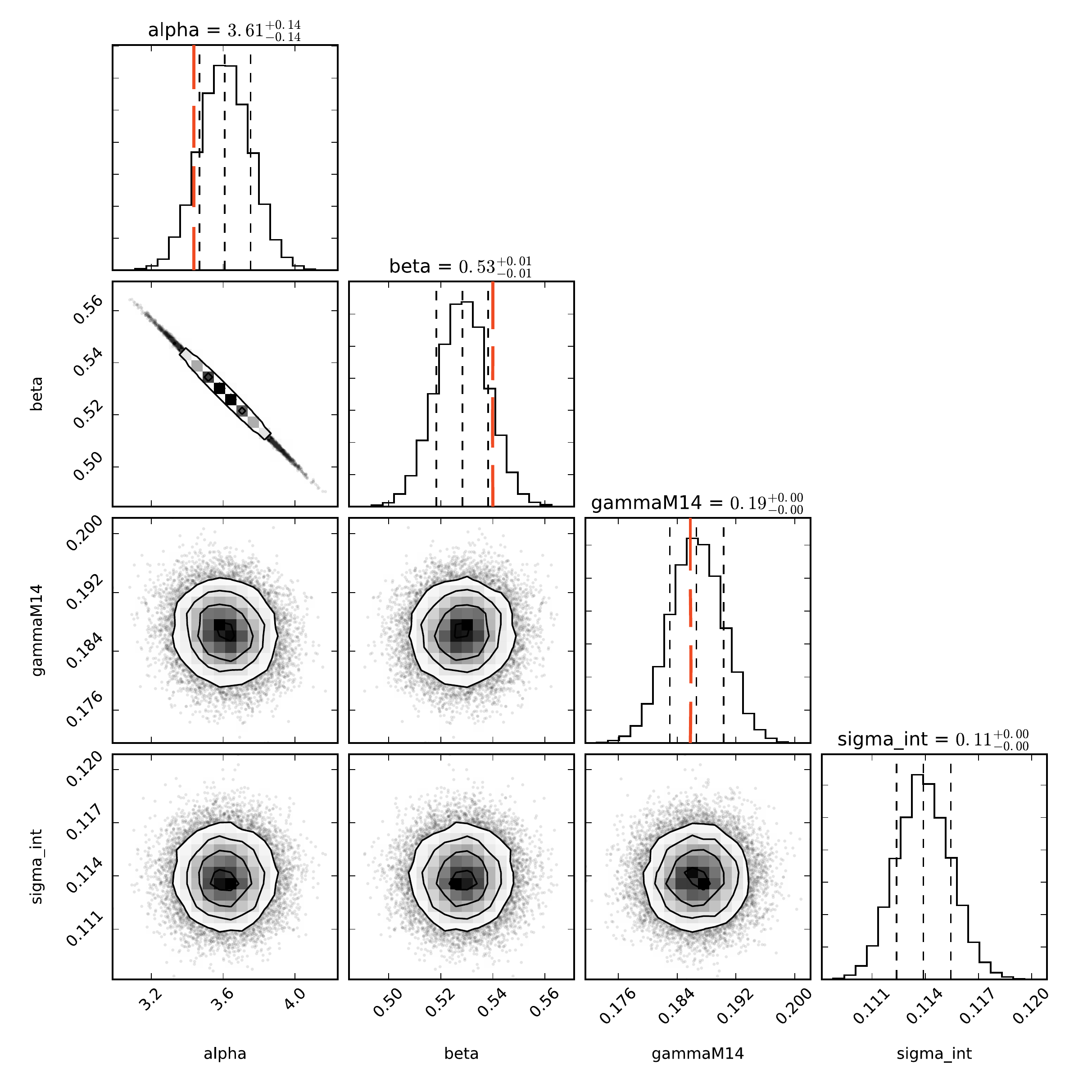}
    \caption{The posterior distribution functions for $\alpha$, $\beta$, $\gamma$, and $\sigma_{int}$ for the \citet{hen12} 3D sample.  The red lines represent estimates for $\alpha$, $\beta$, and $\gamma$ done by binning the stellar mass, halo mass, and magnitude gap and applying a linear fit.  This figure was constructed using M14.  In our analysis of the \citet{hen12} sample, $\gamma$ is significantly non-zero.  Additionally, our Bayesian analysis results agree with the simple linear binned estimates for each parameter.}
    \label{fig:SMHMR-Triangle-T}
\end{figure}

For our 2D Bayesian analysis, we use strong priors on the error distributions, stellar masses, and magnitude gaps, because (a) we can measure the uncertainties and (b) there are no observational uncertainties. In the 3D analysis, we also use strong priors on the halo masses, since we do not use the inferred caustic masses. 

\begin{figure}
    \centering
    \includegraphics[width=8cm]{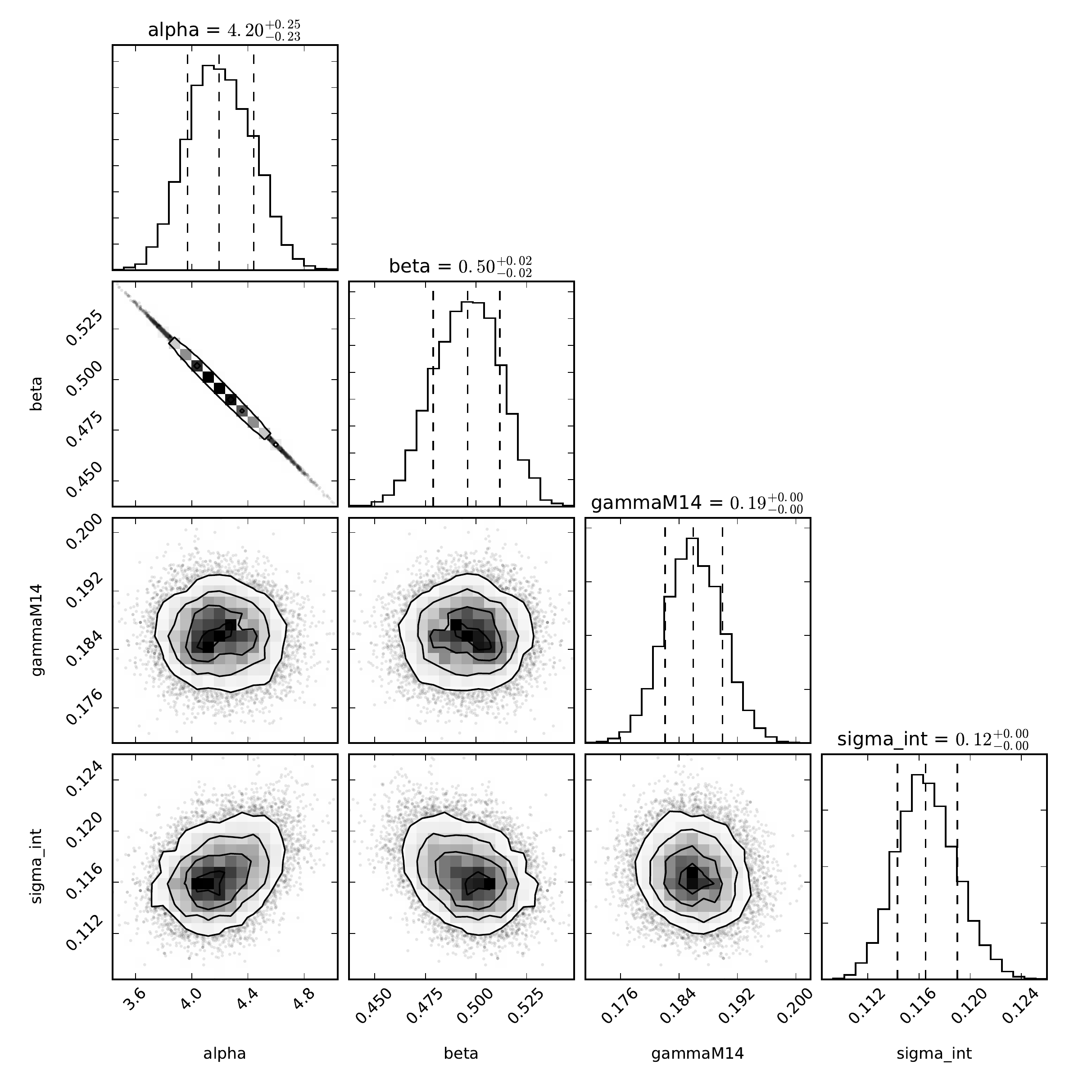}
    \caption{The posterior distribution functions for $\alpha$, $\beta$, $\gamma$, and $\sigma_{int}$ for the \citet{hen12} 2D projected sample.  This figure was constructed using M14.  The posteriors agree with our results for the 3D sample shown in Figure~\ref{fig:SMHMR-Triangle-T}.  Again, we find that $\gamma$ is definitively non-zero.}
    \label{fig:SMHMR-Triangle-P}
\end{figure}

Figures~\ref{fig:SMHMR-Triangle-T} and \ref{fig:SMHMR-Triangle-P} show convergence for each of the four variables.  In these same figures, we see that $\alpha$ and $\beta$ are covariant, which follows since they represent the intercept and slope, respectively.  Of greater importance, neither remaining variable pair is covariant, which emphasizes that neither $\gamma$ or $\sigma_{int}$ are inherently tied to our measurement of the slope.

When we look beyond the distributions and at the values of the posterior distributions shown in Figures~\ref{fig:SMHMR-Triangle-T} and \ref{fig:SMHMR-Triangle-P}, we see that the best fit estimates for $\alpha$, $\beta$, $\gamma$, and $\sigma_{int}$ are within 1 or 2$\sigma$ of one another.  In other words, for these parameters, the model which includes uncertainty can correctly recover the 3D truth, where there is no uncertainty.  The ability to reproduce the results of the 3D model using the 2D projected sample is key because these samples represent the same underlying distribution, but with different uncertainties in each measurement.  Since the posteriors of $\alpha$, $\beta$, $\gamma$, and $\sigma_{int}$ in the 2D projected sample (Figure~\ref{fig:SMHMR-Triangle-P}) agree with those from the 3D sample (Figure~\ref{fig:SMHMR-Triangle-T}), we conclude that if we can accurately account for the uncertainty in stellar mass, halo mass, and magnitude gap, then we can measure the underlying posterior distributions of $\alpha$, $\beta$, $\gamma$, and $\sigma_{int}$ for the SDSS-C4 sample even though we have no 3D dataset.  
 
We also looked at varying the scatter on the magnitude gap by adding a small uncertainty, 0.05, to the the magnitude gaps to account for the fact that a fraction of our 2D light-cone ``observed'' gaps do not exactly agree with the true magnitude gaps.  However, this made no difference to the posterior distribution shown in Figure~\ref{fig:SMHMR-Triangle-P}.  

The most significant result shown in Figures~\ref{fig:SMHMR-Triangle-T} and \ref{fig:SMHMR-Triangle-P} is that $\gamma$ is significantly non-zero.  To highlight this further, we bin the 3D simulated data according to magnitude gap and halo mass and measure the binned stellar mass directly.  From this we can directly (albeit crudely) estimate $\alpha$, $\beta$, and $\gamma$, using linear fits.  The results are shown as the red vertical dashed bars in Figure \ref{fig:SMHMR-Triangle-T}.  This measurement highlights that for the \citet{hen12} prescription of the MILLENNIUM simulation, the magnitude gap is indeed a latent parameter found in the SMHM relation and should be incorporated into other SMHM relations.

Additionally, we note that the quantitative results for the analysis done using M12, shown in Table~\ref{tab:results}, and M14 are well within 2$\sigma$ of one another for both $\alpha$, $\beta$ and $\sigma_{int}$.  However, we find that $\gamma$ is slightly smaller when M12 is used than when M14 is used in our analysis.  We posit that this difference results from the different magnitude gap distributions associated with each respective sample.  

\subsubsection{Model without Magnitude Gap}
\label{subsubsec:nogamma-sims}
To determine the impact of the magnitude gap on $\sigma_{int}$, we ran our Bayesian MCMC model using a linear relation, similar to the formalism used for the high halo mass portion of \citet{yan09} and \citet{mos10, mos13}.  To do this, we replace Equation~\ref{eq:smhm_relation} with:
\begin{equation}
y_i = \mathcal{N}(\alpha + \beta x_{i}, \sigma_{int}^2).
\label{eq:model_nogamma}
\end{equation}
where $\alpha$ and $\beta$ are the intercept and slope and $\sigma_{int}$ is the intrinsic scatter.  For these models, we measured the posterior distributions for $\alpha$, $\beta$, and $\sigma_{int}$.  

Using the same Bayesian formalism and MCMC model described in Section~\ref{sec:model}, but instead using Equation~\ref{eq:model_nogamma}, we find that $\sigma_{int} = 0.159 \pm 0.002$ and $\sigma_{int} = 0.160 \pm 0.003$ for the 3D and 2D catalogs, respectively.  Therefore, comparing this posterior to those obtained when incorporating the magnitude gap, $\sigma_{int} = 0.114 \pm 0.002$ and $0.117 \pm 0.002$ (3D and 2D), we find that the intrinsic scatter significantly decreases due to the inclusion of the magnitude gap as a third parameter because the two values differ by greater than 3$\sigma$.  Additionally, incorporating the magnitude gap reduces $\sigma_{int}$ by by 28.3\% in the 3D sample and 26.9\% in the 2D projected sample.  We also observe that the inclusion of the magnitude gap does not impact the measurement of the slope, $\beta$, because the posterior values determined when accounting for the magnitude gap and when not agree.

\subsection{SDSS-C4 Sample}\label{subsec:R-C4}
\subsubsection{SMHM Relation}\label{subsubsec:R-SMHM}
\begin{figure}
    \centering
    \includegraphics[width=8cm]{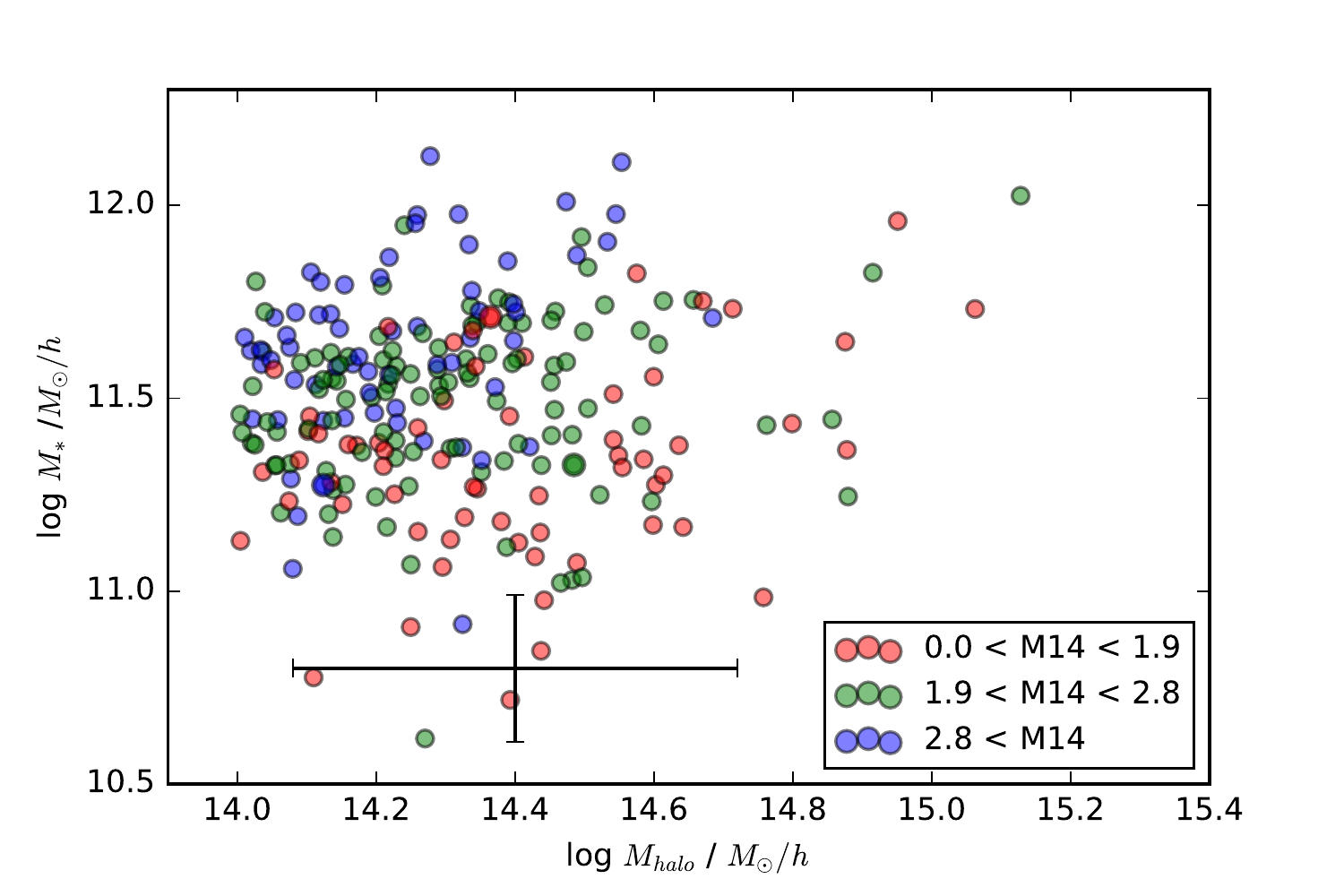}
    \caption{The SMHM relation for the SDSS-C4 sample of clusters binned via M14 measurements.  The black bars represent the median error bars in caustic halo mass, 0.32 dex, and the error in stellar mass, 0.19 dex.  Similar to the results of the \citet{hen12} 2D projected data, shown in Figure~\ref{fig:SMHMR-HenP}, a magnitude gap -- stellar mass stratification exists in the real universe, where measurement errors are found on all three parameters shown.}
    \label{fig:SMHMR-C4-binned-M14}
\end{figure}

\begin{figure}
    \centering
    \includegraphics[width=8cm]{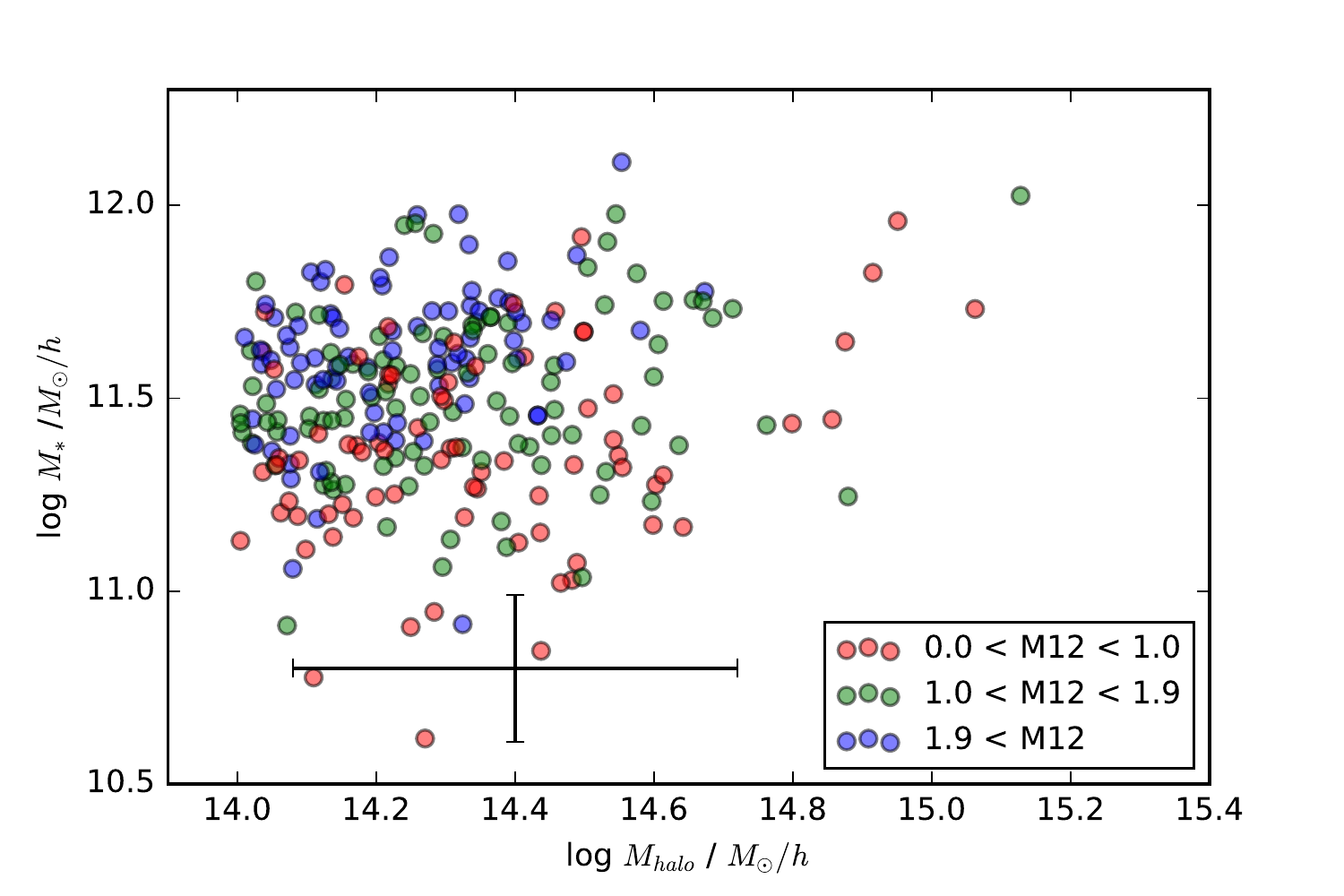}
    \caption{The SMHM relation for the SDSS-C4 sample of clusters binned using the M12 measurements.  The black bars represent the median error bars in halo mass, 0.32 dex, and stellar mass, 0.19 dex.  When compared to Figure~\ref{fig:SMHMR-C4-binned-M14}, this figure highlights that a similar magnitude gap -- stellar mass stratification is observed regardless of the choice of n$^{th}$ brightest cluster member.}
    \label{fig:SMHMR-C4-binned-M12}
\end{figure}

In this section, we present the qualitative results from our analysis of the SDSS-C4 samples using both M14 and M12.  The SMHM relations shown in Figures~\ref{fig:SMHMR-C4-binned-M14} and \ref{fig:SMHMR-C4-binned-M12} plot the stellar masses estimated using the \citet{bel03} M/L ratio relation against our caustic halo masses and utilize three non-uniform magnitude gap bins to represent the clusters with high (blue), intermediate (green), and low (red) magnitude gaps.

We use three color bins, unlike in Figures~\ref{fig:SMHMR-HenT} and~\ref{fig:SMHMR-HenP}, because a noisier color gradient exists in our SDSS-C4 observations.  This additional noise is primarily caused by the photometric measurement error associated with the corrected BCG magnitudes for our SDSS-C4 observations.  Additionally, the relative lack of redshift information, compared to the \citet{hen12} data, results in our observed M14 and M12 values being more sensitive to our red sequence fit.  The lack of clusters, 2700 in the simulations and 236 (M14) or 254 (M12) in the SDSS-C4 sample, also makes the gradient more difficult to clearly observe.  Thus, binning the data by magnitude gap allows us to more easily convey the impact of incorporating the magnitude gap and the observed stratification.  We chose non-uniform bin widths to emphasize the similarity between the M14 and M12 samples.  Furthermore, a comparison between Figures~\ref{fig:SMHMR-C4-binned-M14} and \ref{fig:SMHMR-C4-binned-M12} suggests that this stratification may be independent of the choice of n$^{th}$ brightest cluster member used to measure the magnitude gap.

By binning our data, Figures~\ref{fig:SMHMR-C4-binned-M14} and \ref{fig:SMHMR-C4-binned-M12} illustrate that a magnitude gap stratification with some scatter exists in the real universe.  For a fixed halo mass, as the stellar mass increases, the color of the points transitions from red to green to blue, with some additional scatter, which represents that stellar mass and magnitude gap are positively correlated.  

Just as for our analysis of the \citet{hen12} prescription of the MILLENNIUM simulation, we clearly observe a trend relating magnitude gap and stellar mass at fixed halo mass.  Furthermore, Figures~\ref{fig:SMHMR-C4-binned-M14} and~\ref{fig:SMHMR-C4-binned-M12} show that, like in the simulations, a bifurcation between high and low magnitude gap clusters (fossils and non-fossils) is an oversimplification.  Instead, we treat this relationship as a stratification, which is independent of the optical and X-ray definitions of fossil galaxies, and instead extends to all clusters because at fixed halo mass, as the magnitude gap between the BCG and any selected n$^{th}$ brightest member increases, on average, the stellar mass of the BCG similarly increases.  Thus, allowing the magnitude gap to be treated as a relative proxy for the stellar mass of a BCG. 

Additionally, we highlight the accuracy of our magnitude gap measurements by looking at how the distributions of the magnitude gap measured in our SDSS-C4 data compare to those measured in our 2D \citet{hen12} projected data.   
\begin{figure}
    \centering
    \includegraphics[width=8cm]{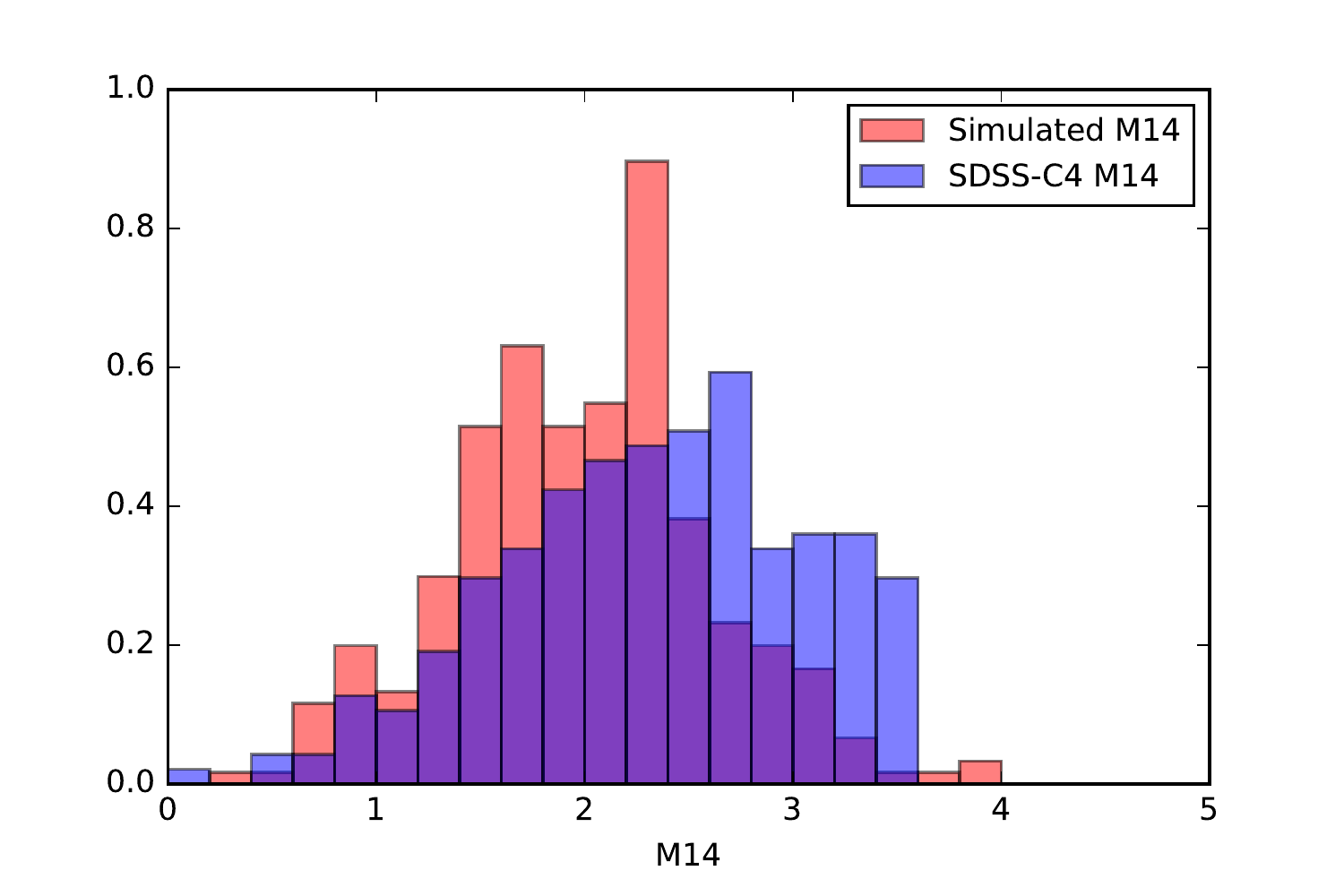}
    \caption{The distribution for the SDSS-C4 M14 values and the distribution for the 2D projected \citet{hen12} M14 values.  The similarity highlights that our 4th brightest SDSS-C4 cluster members are accurately identified using the red sequence and available redshift information.}
    \label{fig:Sims-C4-M14}
\end{figure}
\begin{figure}
    \centering
    \includegraphics[width=8cm]{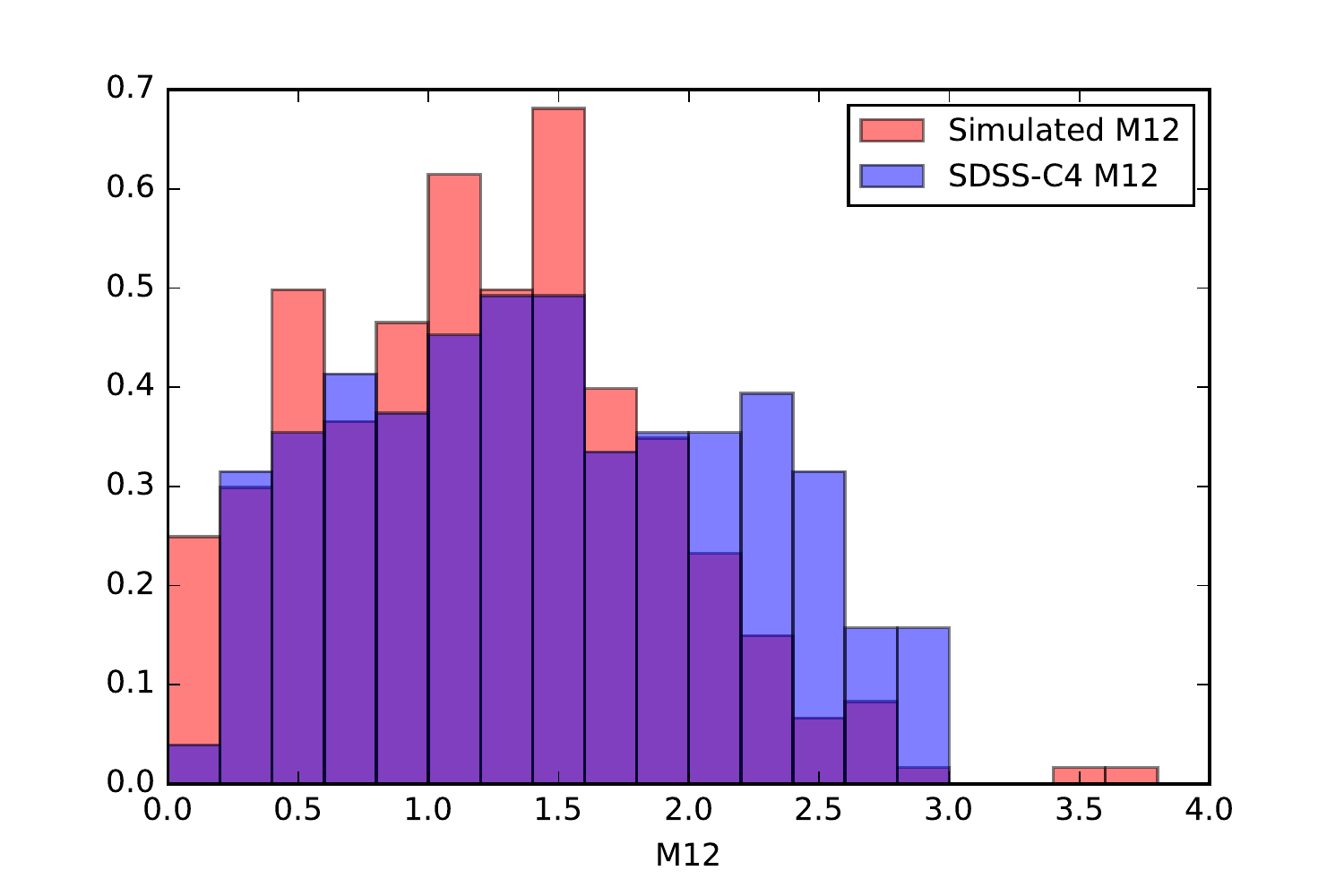}
    \caption{The distribution for the SDSS-C4 M12 values and the distribution for the 2D projected \citet{hen12} M12 values.  A comparison illustrates that our 2nd brightest SDSS-C4 cluster members are accurately identified using available redshift information and red sequence fitting.}
    \label{fig:Sims-C4-M12}
\end{figure}
In Figures~\ref{fig:Sims-C4-M14} and \ref{fig:Sims-C4-M12}, we have normalized the distributions for comparison, which results in a y-axis that represents a relative number.  We find that the overall shape of the distributions for the SDSS-C4 data (in blue) and the \citet{hen12} data (in red) are quite similar.  The primary difference is that the SDSS-C4 magnitude gap values are slightly larger, which may result from projection effects, due to the lack of available redshift information or our red sequence fitting.  

\subsubsection{Quantitative Impact}
\label{subsubsec:R-QI}
As done in Section~\ref{subsubsec:MILL-QI}, we evaluate the impact of incorporating the magnitude gap into our SMHM relation using the previously described (Section~\ref{sec:model}) MCMC model, Bayesian formalism, and linear SMHM relation (Equation~\ref{eq:smhm_relation}).  The model used for our SDSS-C4 sample differs slightly from that used for the simulated data because of minor differences related to our estimation of the uncertainties in the measurements of stellar mass, halo mass, and magnitude gap.  For halo mass, we used the relation between the number of galaxies used to construct the caustic phase space and the measurement uncertainty presented in \citet{gif13a}, while for the 2D simulated light-cone data, we used a fixed error of 0.35 dex, since it has a deep magnitude limit.  For the stellar mass, we increased the measurement uncertainty from 0.03 dex (sims) to 0.19 dex.  We reached 0.19 dex by assuming a 0.1 dex error in the M/L ratio \citep{bel03} and combining it with the 0.1 magnitude precision in both the BCG's r and i-band magnitudes used to determine the color used in the \citet{bel03} relation.  For the magnitude gap, due to the BCG photometry, we also assumed an uncertainty of 0.1 magnitudes, in contrast to 0.0 magnitudes used in the simulations.  Since we use the SDSS Petrosian magnitudes for our 4th and 2nd brightest galaxies, we assume that the measurement uncertainties associated with those magnitudes are negligible.  Additionally, for each uncertainty, we add a random $\beta$ distribution term to our error estimates to encapsulate the uncertainty associated with each one of our error estimates, given by Equations~\ref{eq:beta_y}, \ref{eq:beta_x}, and \ref{eq:beta_z}.  

\begin{figure}
    \centering
    \includegraphics[width=8cm]{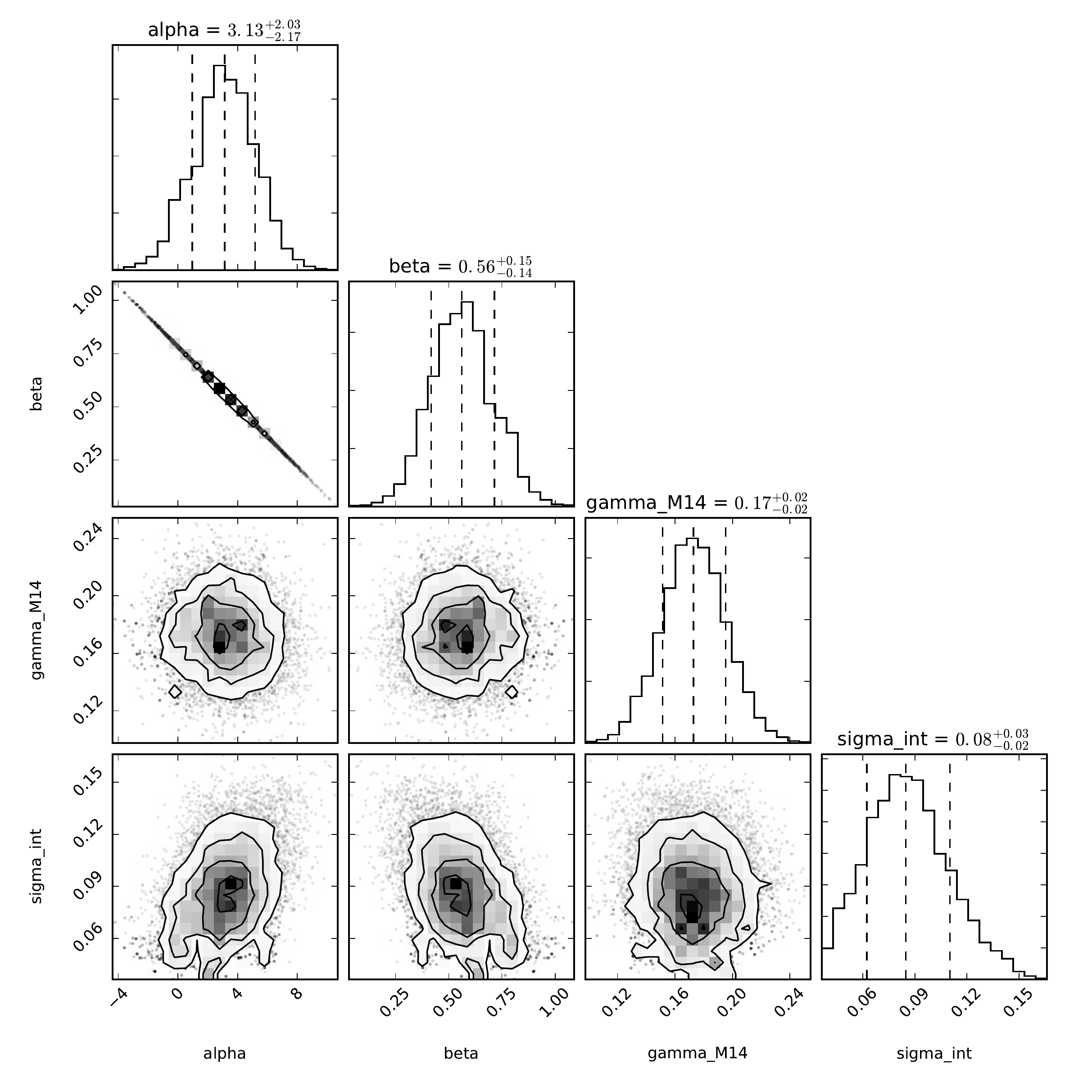}
    \caption{The posterior distributions for $\alpha$, $\beta$, $\gamma$, and $\sigma_{int}$ for the SDSS-C4 sample measured using M14. Like in the \citet{hen12} 2D projected sample, shown in Figure~\ref{fig:SMHMR-Triangle-P}, we see that $\gamma$ is definitely non-zero in the real universe.  Additionally, we find that $\sigma_{int}$ is below 0.1 dex.}
    \label{fig:SMHMR-C4-Triangle-M14}
\end{figure}

\begin{figure}
    \centering
    \includegraphics[width=8cm]{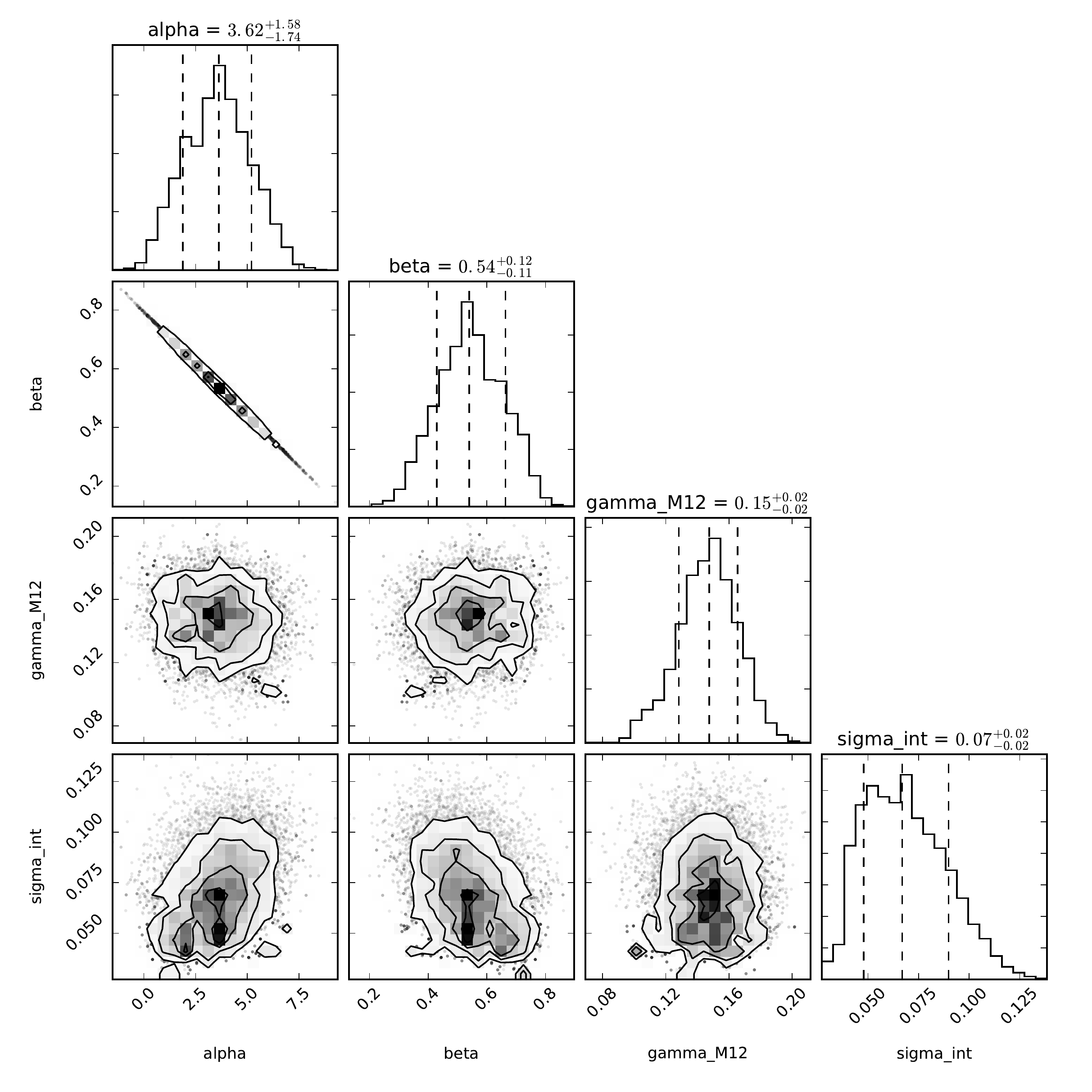}
    \caption{The posterior distributions for $\alpha$, $\beta$, $\gamma$, and $\sigma_{int}$ for the SDSS-C4 sample measured using M12.  The posteriors agree with the results shown in Figure~\ref{fig:SMHMR-C4-Triangle-M14}.  Similarly, $\gamma$ is significantly non-zero, and $\sigma_{int}$ is well below 0.1 dex.}
    \label{fig:SMHMR-C4-Triangle-M12}
\end{figure}

Using this approach, we present triangle plots, shown in Figures~\ref{fig:SMHMR-C4-Triangle-M14} and \ref{fig:SMHMR-C4-Triangle-M12}, which show the distributions of $\alpha$, $\beta$, $\gamma$, and $\sigma_{int}$ for both the SDSS-C4 M14 and M12 samples.  Each plot was generated after 10 million steps including an approximate 2 million step burn in.  

Figures~\ref{fig:SMHMR-C4-Triangle-M14} and~\ref{fig:SMHMR-C4-Triangle-M12} show the marginalized 1D or 2D posteriors after convergence.  We see that only $\alpha$ and $\beta$ are covariant, as was the case for our simulations and is expected in a linear regression.  Additionally, the values based on the posteriors for $\alpha$, $\beta$, $\gamma$, and $\sigma_{int}$ when determined using M14 are within 1$\sigma$ of their counterparts done using M12 (see Table \ref{tab:results}).  Thus, the choice of n$^{th}$ brightest galaxy used to measure the magnitude gap appears to have little impact on the measured parameters, which strengthens our argument that the magnitude gap-stellar mass stratification is not dependent on our choice of n$^{th}$ brightest cluster member, based on Figures~\ref{fig:SMHMR-C4-binned-M14} and \ref{fig:SMHMR-C4-binned-M12}.  We note that the posterior constructed using M12 leads to a slightly lower value of $\gamma$ as observed in the \citet{hen12} simulated data; however, since the error bars are larger on the posteriors of our SDSS-C4 data than on the posteriors of the \citet{hen12} data, these $\gamma$ posteriors are in agreement. 

The primary result from Figures~\ref{fig:SMHMR-C4-Triangle-M14} and \ref{fig:SMHMR-C4-Triangle-M12} is that $\gamma$ is definitively non-zero in our SDSS-C4 sample, as we observed in the \citet{hen12} simulation.  This observation highlights that we must treat the magnitude gap as a latent third parameter in the SMHM relation. Additionally, we note that for the first time, the observational estimate for the intrinsic scatter has moved below a precision of 0.1 dex.  We discuss the implications of these results further in Section~\ref{sec:Discussion}.

Recall that for the 2D simulated data, we varied the scatter of the magnitude gap but found no difference in the results.  We investigate this again in the SDSS-C4 data.  In the nominal analysis, we assume that the magnitude gaps have a measurement error of 0.1 magnitudes.  As described earlier, we include an additional stochastic uncertainty on this error using a beta distribution, which adds as much as $\pm 0.06$ magnitudes to the gap error.  As we push this up to larger magnitude gap errors we begin to see the stretch factor $\gamma$ changing its median in the posterior at $\sigma_{z_{i}}(M14) = 0.4$. However, given that we have used all available spectroscopic information for our clusters, have well-determined red-sequence membership, and that we have re-created the gap measurement distribution in a realistic mock sky, we are confident that our gap measurements are as accurate as we describe (0.1 magnitudes).  Additionally, we also examine the sensitivity of our results on our measurement uncertainties for stellar mass and halo mass.  We find that if we vary the uncertainty in halo mass by 25\% the results of our posterior are all in agreement with our results presented in Table~\ref{tab:results}.  When we vary the uncertainty in stellar mass by 25\%, we see the same trend between slope and error measurement presented in Figure 10 of \citet{tin16}.  Additionally, all of the parameters, except $\sigma_{int}$ are in agreement with their values in Table~\ref{tab:results}.  The discrepancy for $\sigma_{int}$ is expected due to the relationship between error in stellar mass and intrinsic scatter.  We further justify our choice of 0.19 dex for the measurement error in stellar mass in Sections~\ref{subsubsec:nogamma-C4} and \ref{subsec:Impact}.    

\subsubsection{Model without incorporating the Magnitude Gap}
\label{subsubsec:nogamma-C4}
\begin{figure}
    \centering
    \includegraphics[width=8cm]{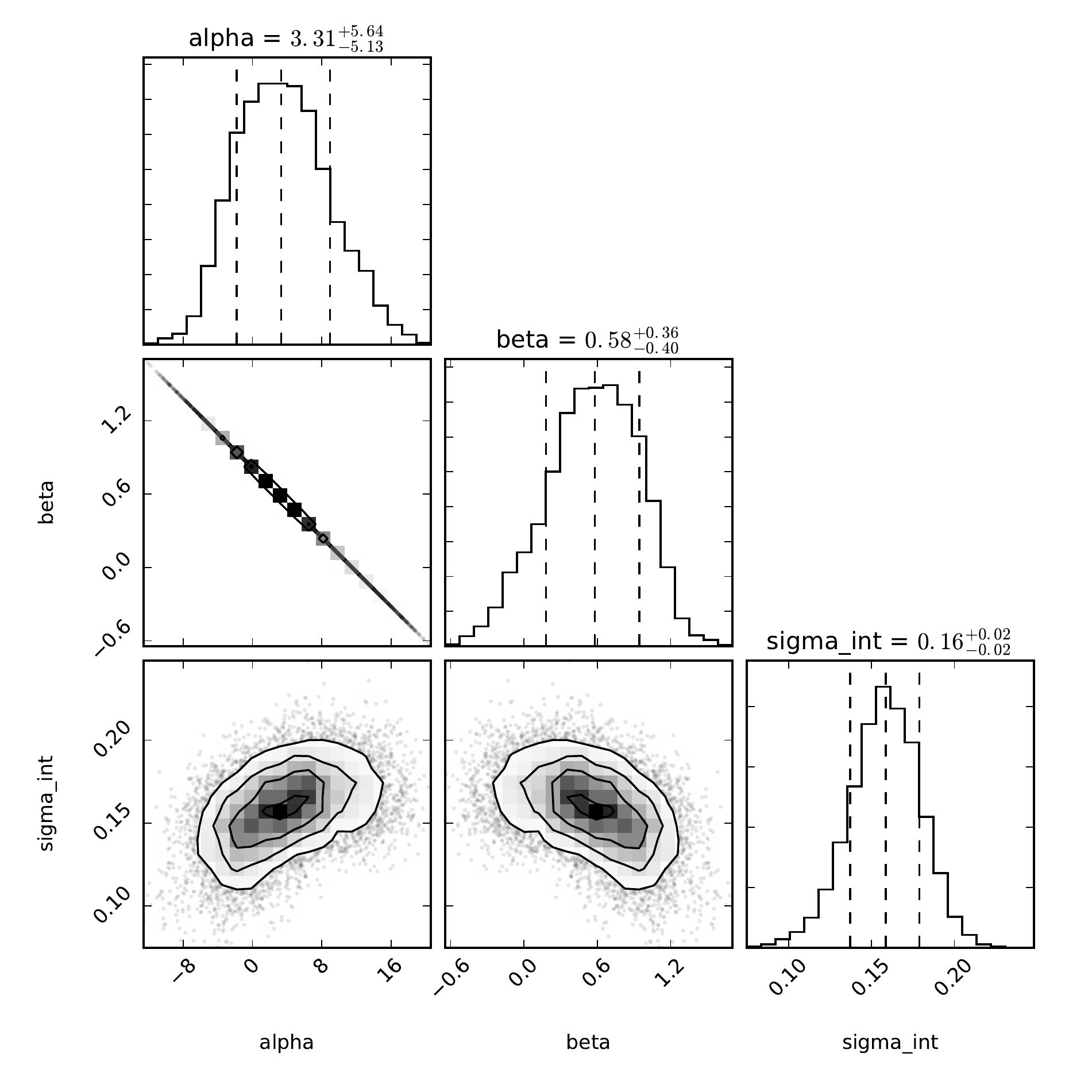}
    \caption{The posterior distributions for $\alpha$, $\beta$, and $\sigma_{int}$ for the SDSS-C4 sample measured without incorporating the magnitude gap.  A comparison to Figures~\ref{fig:SMHMR-C4-Triangle-M14} and \ref{fig:SMHMR-C4-Triangle-M12} highlights that $\beta$ is unchanged when incorporating the magnitude gap; however $\sigma_{int}$ is significantly higher when the magnitude gap is unaccounted for.}
    \label{fig:SMHMR-C4-Triangle-nogamma}
\end{figure}

To determine the impact of incorporating the magnitude gap on $\sigma_{int}$ for the observed SMHM relation, we again use the MCMC model and Bayesian formalism, but instead use the linear relation, Equation~\ref{eq:model_nogamma}, presented in Section~\ref{subsubsec:nogamma-sims}, which does not incorporate the magnitude gap.  The results of this analysis are shown via a triangle plot in  Figure~\ref{fig:SMHMR-C4-Triangle-nogamma}, 
which compares the posterior distributions of $\alpha$, $\beta$, and $\sigma_{int}$.  As with our previous MCMC runs, we see a convergence and find that only $\alpha$ and $\beta$ are covariant, while $\sigma_{int}$ does not depend on either parameter.  When comparing the results of this analysis to Figures~\ref{fig:SMHMR-C4-Triangle-M14} and \ref{fig:SMHMR-C4-Triangle-M12} we find that the values of the slope, $\beta$, are in agreement with one another.

The primary reason we ran an analysis using the model without the stretch parameter (Equation~\ref{eq:model_nogamma}) was to determine the impact of the magnitude gap on $\sigma_{int}$.  Of note, the $\sigma_{int}$ we measure using our 2 component linear model, 0.159 $\pm$ 0.021, is in excellent agreement with the best prior estimates of the intrinsic scatter from both observations and simulations for the high mass portion of the SMHM relation, 0.15 dex \citep{pil17}, 0.16 dex \citep{tin16}, and 0.17 dex \citep{kra14}, which highlights the consistency of our measurements with other studies.  When we compare the results of our model without the magnitude gap to those presented in Section~\ref{subsubsec:R-QI} which incorporate it, we find that including the magnitude gap significantly reduces the the median value of $\sigma_{int}$ by either 0.074 or 0.092 dex.  Thus, incorporating the magnitude gap leads to either a 45\% or 58\% decrease in the intrinsic scatter.  Additionally, we note that the values of the intrinsic scatter obtained when incorporating the magnitude gap are not within one sigma of the estimate obtained when not incorporating it.  Thus, the magnitude gap is a latent parameter in the SMHM relation that significantly reduces the intrinsic scatter in this relation and allows us to enter the realm of sub 0.1 dex precision.  

For comparison, we find that the reduction of $\sigma_{int}$ is far greater for the SDSS-C4 data than what we find in the \citet{hen12} prescription of the MILLENNIUM simulation.  This may result from the error bars for our posterior distributions being larger for our SDSS-C4 observations than for the simulations.  Alternatively, unlike in our SDSS-C4 observations, where we assumed that the uncertainty in the stellar mass is 0.19 dex, for the 3D \citet{hen12} prescription of the MILLENNIUM simulation, we assumed that the stellar mass had no measurement uncertainty associated with it.  However, this is likely an underapproximation, because there is likely some additional uncertainty associated with those stellar mass estimates.

In addition to the reduction in the inferred intrinsic scatter, we find that the size of the error bars on the posterior distribution of the slope, $\beta$, significantly decreases when the magnitude gap is incorporated.  As shown in Figures~\ref{fig:SMHMR-C4-Triangle-M14} and~\ref{fig:SMHMR-C4-Triangle-M12}, $\beta$ has error bars of approximately 0.15 and 0.11 dex respectively.  In contrast, when we don't incorporate the magnitude gap, the error bars on $\beta$ increase to between 0.36 and 0.40 dex, as shown in Figure~\ref{fig:SMHMR-C4-Triangle-nogamma}.  Thus, the uncertainty associated with our slope decreases by between 0.21 to 0.29 dex, when we include the magnitude gap in our analysis.  This decrease likely occurs because for a fixed magnitude gap we use fewer points spread over a smaller range in stellar mass to fit the slope, making it more tightly constrained.  Therefore, incorporating the magnitude gap not only allows us to significantly reduce the intrinsic scatter in the stellar mass at fixed halo mass, it also allows us to reduce the uncertainty on our measurement of the slope.  

To verify that the three parameter model (given by Equation~\ref{eq:smhm_relation}) does indeed reproduce the data better than the two parameter model (given by Equation~\ref{eq:model_nogamma}), we use the posterior predictive distribution to compare the values of the stellar mass and magnitude gap given by each of the models.  For comparison, we measure the R coefficient to be R=0.428 for our observed SDSS-C4 data.  When averaged over the MCMC trace in our Bayesian model, we find that the two parameter model yields an $\langle R \rangle=0.169$, while the three parameter model yields $\langle R \rangle=0.410$, which allows us to conclude that the 2 parameter model does not reproduce the observed correlation between stellar mass and magnitude gap that is observed in our data.   

Additionally, to further verify that the reduction of $\sigma_{int}$ results from incorporating the magnitude gap, and not just the inclusion of a randomly selected third parameter, we reran our Bayesian analysis using randomized values of M14 in Equation~\ref{eq:smhm_relation}.  Doing so removes the correlation between M14 and stellar mass, and results in a posterior distribution with a $\gamma$ that is equal to $0.0 \pm 0.02$.  Furthermore, the other parameters contained in this posterior, including the measurement of $\sigma_{int}$, agree with the values presented in Figure~\ref{fig:SMHMR-C4-Triangle-nogamma}, which further highlights that the reduction in $\sigma_{int}$ and measurement of non-zero $\gamma$ result because stellar mass and M14 are correlated due to the hierarchical growth of the BCG.  Therefore, it is the incorporation of this specific third parameter, the magnitude gap, and not any random third parameter, that leads to the significant reduction in $\sigma_{int}$.

\begin{deluxetable*}{cccccc}
	\tablecaption{Posterior Distribution Results}
	\tablecolumns{6}
	\tablewidth{0pt}
	\tablehead{\colhead{Data} & 
	\colhead{Magnitude Gap} & 
	\colhead{$\alpha$} &
	\colhead{$\beta$} &
	\colhead{$\gamma$} &
	\colhead{$\sigma_{int}$}} 
\startdata
\citet{hen12}-3D & Not Incorporated & 4.31 $\pm$ 0.20 & 0.51 $\pm$ 0.01 & & 0.159 $\pm$ 0.002 \\
\citet{hen12}-2D & Not Incorporated & 4.04 $\pm$ 0.27 & 0.53 $\pm$ 0.02 & & 0.160 $\pm$ 0.003 \\
\citet{hen12}-3D & M14 & 3.61 $\pm$ 0.14 & 0.53 $\pm$ 0.01 & 0.187 $\pm$ 0.004 & 0.114 $\pm$ 0.002 \\
\citet{hen12}-2D & M14 & 4.19 $\pm$ 0.23 & 0.50 $\pm$ 0.02 & 0.186 $\pm$ 0.004 & 0.117 $\pm$ 0.002 \\
SDSS-C4 & M14 & 3.13 $\pm$ 2.09 & 0.56 $\pm$ 0.15 & 0.173 $\pm$ 0.022 & 0.085 $\pm$ 0.024 \\
SDSS-C4 & Not Incorporated & $3.31\substack{+5.64 \\ -5.13}$\ & $0.58\substack{+0.36 \\ -0.40}$\ &  & 0.159 $\pm$ 0.021\\ 
\citet{hen12}-3D & M12 & 4.07 $\pm$ 0.15 & 0.51 $\pm$ 0.01 & 0.147 $\pm$ 0.004 & 0.124 $\pm$ 0.002 \\
\citet{hen12}-2D & M12 & 3.66 $\pm$ 0.24 & 0.55 $\pm$ 0.02 & 0.158 $\pm$ 0.004 & 0.123 $\pm$ 0.003 \\
SDSS-C4 & M12 & $3.62\substack{+1.58 \\ -1.74}$\ & $0.54\substack{+0.12 \\ -0.11}$\ & 0.147 $\pm$ 0.019 & 0.067 $\pm$ 0.020 \\
SDSS-C4, Abundance Matching & Not Incorporated & -0.33 $\pm$ 0.15 & 0.84 $\pm$ 0.01 & & \\

\enddata
\label{tab:results}
\end{deluxetable*}

\section{Discussion}
\label{sec:Discussion}
\subsection{Comparisons to the Literature}
\label{subsec:Complit}

We have presented results that show that in both a simulated semi-analytic model of a low-redshift universe and in the observed universe, there is a stratification in stellar mass with the magnitude gap at fixed halo mass for cluster-sized halos. This result is found at high confidence, such that the measured stretch factor $\gamma$ is many standard deviations away from zero. The inclusion of the magnitude gap as a latent parameter in the cluster-scale SMHM relation reduces the scatter $\Delta(\rm{log} M_{*}|\rm{log} M_{halo})$ by a significant amount.  At the same time, it also reduces the error bar on the inferred slope, $\beta$.

The physical importance of the detection of the magnitude gap as a latent variable (albeit an observational one) directly relates to the BCG growth history.  This history is built into the \citet{hen12} and \citet{guo10} prescriptions of the MILLENNIUM simulation, in which BCGs grow hierarchically via major and minor mergers \citep{del07}.  The more massive BCGs are those that have undergone more mergers, and in agreement with \citet{solanes16}, have the largest magnitude gaps.  Therefore, the observation of a magnitude gap stratification in the SDSS-C4 data acts as observational evidence that in the real universe, BCGs predominantly grow hierarchically.  Additionally, we posit that the magnitude gap may be able to trace the assembly history of both the BCG and the cluster.

As we have shown, in agreement with \citet{solanes16}, the clusters with the largest magnitude gaps and stellar masses have likely undergone more mergers than small gap, low magnitude gap clusters.  If a similar dynamical timescale for the mergers that occur in each cluster exists, then the clusters with the largest stellar mass and magnitude gaps would likely be the clusters that formed first.  This hypothesis is supported by Figure 7 of \citet{mat17}.  Using the hydronamical EAGLE simulation, \citet{mat17} construct a stellar mass halo mass relation over the halo mass range of $11.0 < log10(M_{200,DMO}) < 14.5$, in which individual clusters are color coded by the redshift when half of the halo mass was assembled.  This SMHM relation agrees with our previously stated assumption, and shows that a stratification between the stellar mass and formation time exists in the EAGLE simulation; at a fixed halo mass, the most massive galaxies are found in the halos that form at the earliest redshift \citep{mat17}.  This stratification appears to exist, but is difficult to analyze at halo masses close to our range of interest because the EAGLE simulation does not contain enough high halo mass clusters.  Thus, a direct comparison between our studies would be difficult at this time because we can not yet determine how and if the magnitude gap stratification relates to the formation redshift stratification and if the magnitude gap accurately scales with the formation redshift to trace the assembly of the cluster.

\begin{figure}
    \centering
    \includegraphics[width=8cm]{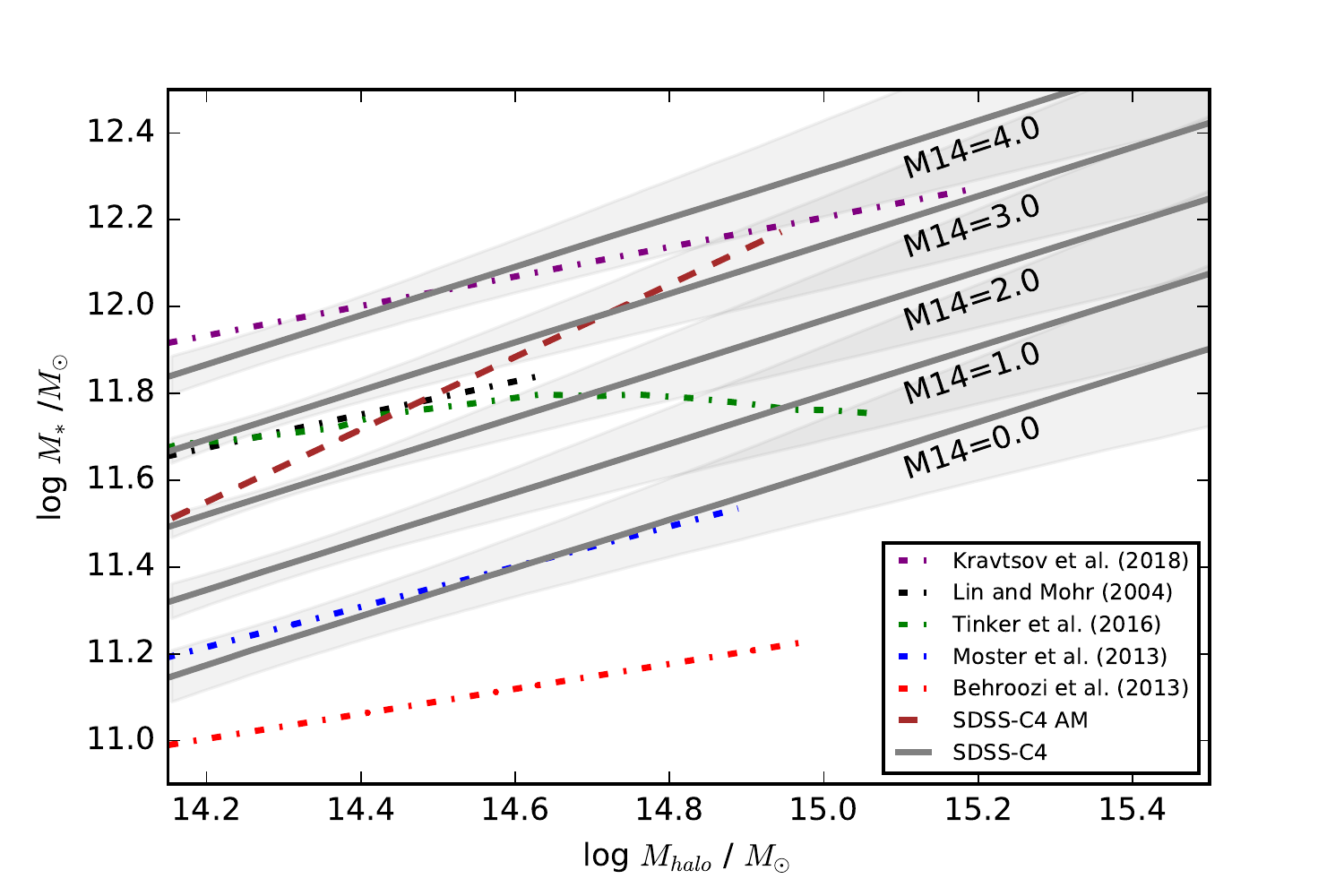}
    \caption{The dot-dashed lines are 5 different previously published SMHM relations, from \citet{beh13}, \citet{mos13}, \citet{tin16}, \citet{lin04}, and \citet{kra14}.  The dashed brown line represents our SDSS-C4 relation in which halo masses were estimated via abundance matching.  The solid gray lines represent the best fit for our SMHM relation done using M14 for five different values of M14.  The shaded region represents the 1$\sigma$ region surrounding each of the 5 magnitude gap fits.  Each of the stellar masses for this figure is scaled to a Salpeter IMF.  The gray lines highlight that incorporating the magnitude gap can account for as much as 0.7 dex in stellar mass and may account for the majority of offsets between previously reported SMHM relations.}
    \label{fig:SMHMR-C4-Comparison}
\end{figure}

Although it remains uncertain whether the magnitude gap traces the assembly history of the halo, the discovery of the stellar mass - magnitude gap stratification does solve a pragmatic issue that has existed for the observed cluster-scale SMHM relation.  As discussed in Section~\ref{sec:intro}, a large discrepancy exists between published SMHM relations in the high mass regime, highlighted by Figure 10 in \citet{tin16}.  This discrepancy includes both purely observational studies (like ours and \citet{kra14}), and those that require a strong theoretical prior from the use of abundance matching models \citep[e.g.,][]{beh10,mos10,beh13,mos13,tin16}.  For a fixed halo mass, the estimates for the stellar mass differ by as much as $0.5 - 1.0$ dex.  In Figure \ref{fig:SMHMR-C4-Comparison}, we show numerous SMHM relations from the literature after normalizing to a Salpeter IMF.  On this figure we also highlight our model SMHM relations (plural), in gray, given a specific magnitude gap, where the light gray band defines the 1$\sigma$ error bar from the posterior.  By varying the average magnitude gap (M14) for a sample from 0.0 to 4.0 (the range covered in Figure \ref{fig:Comp-M14}), the stellar masses can vary by as as much as 0.7 dex at fixed halo mass.  Thus the gap as a latent variable can explain the majority of the offsets in the published SMHM relations, assuming those published results used samples with different average magnitude gaps. 

Based on our model, across the top of our SMHM relation are the largest gap systems, i.e., those classified as fossil galaxies \citep{har12}.  The \citet{kra14} sample is small and only a half dozen clusters overlap with our data where we can verify the magnitude gap (using identical data and techniques) to be $\langle M14 \rangle \sim 2.5$, which would classify the sample as representative of fossils.  \citet{kra14} also used ``total'' magnitudes as opposed to our Petrosian magnitudes, thus further increasing the magnitude gap \citep{graham2005}. Moving down in stellar mass in Figure \ref{fig:SMHMR-C4-Comparison}, consider the cluster sample used in \citet{lin04}, which like the \citet{kra14} sample, is based on X-ray selection.  In this case, 14 clusters overlap with our sample and we measure $\langle M14 \rangle \sim 2$, which is indeed lower than the \citet{kra14} average.  Across the bottom end of the relation are the systems with small values of magnitude gaps.  The lower two literature relations in Figure \ref{fig:SMHMR-C4-Comparison} are not based on cluster identifications at all, but simply use galaxy samples.  These samples are sorted according to inferred stellar mass and matched to a simulation.  In other words, all intrinsically bright galaxies are included in the cluster-scale \citet{beh13} and \citet{mos13} analyses, including those which do not reside in cluster-scale environments.  Because of the lack of the use of confirmed clusters in their samples, it would not be surprising that the average magnitude gaps of the centrals in halos $\ge 10^{14}M_{\odot}$ are small.  Our relation does not intersect with the results from \citet{beh13} because \citet{beh13} use SDSS Petrosian magnitudes which have not been corrected for the systematic background subtraction errors.

It is difficult to make a fair comparison of the literature SMHM relations for numerous reasons noted earlier. However, in the context of our model, the above exercise provides a reasonable explanation for the large amplitude variations in the SMHM seen in the literature. Such a variation would be expected if the mean magnitude gap of the observed samples is not held fixed.

Many other works \citep[e.g.,][]{beh10, mos10, beh13, mos13, tin16} rely upon abundance matching to measure the SMHM relation.  The one-to-one matching of the BCGs with the largest stellar mass to the largest values of cluster masses from simulations pays no regard to the magnitude gap and results in the loss of information about the BCG's growth history.  At the same time, we show that ignoring the magnitude gap results in a large increase in the inferred intrinsic scatter (see Section \ref{subsubsec:nogamma-C4}).  In Figure \ref{fig:SMHMR-C4-Comparison}, we plot our SMHM relation for the SDSS-C4 BCGs based on abundance matching to the \citet{hen12} light-cone halo catalog.  The fit lies within our expectations for the full model, but with a steeper slope.

\subsection{Impact on Galaxy Formation Models}
\label{subsec:Impact}
At lower halo masses we expect in-situ star formation, the accretion of gas, as well as stellar and/or AGN feedback to play some role in the growth of stellar mass over time since $z \sim 2$.  While at the highest halo masses, we expect all of these processes to have finished by $z=2$, leaving only hierarchical growth as the dominant mechanism to increase a galaxy's stellar mass since $z \sim 2$.  However, the reported observed intrinsic scatter is nearly constant with halo mass at around $\sigma (M_{*}) < 0.2$ dex \citep{gu2016}. 

\citet{gu2016} used abundance matching and investigated the origin of scatter at fixed halo mass by following the hierarchical buildup of both dark and stellar mass in simulations.  \citet{gu2016} concluded from their model and simulations that there should be a strong mass-dependent scatter in the SMHM relation.  Similarly, this conclusion was also reached using the hydrodynamical EAGLE simulation, where \citet{mat17} estimate the intrinsic scatter in stellar mass at fixed halo mass using different parametric fits of the true and predicted stellar masses to the halo mass and find that the intrinsic scatter decreases, by approximately 0.1 dex, as the halo increases over the range of $11.0 < log10(M_{h}) < 13.0$.
Assuming that the intrinsic scatter measurements reported in previous SMHM relations are dominated by lower mass halos ($log10(M_{h}) < 14.0$), our work provides the first observational evidence that the scatter in stellar mass decreases significantly for centrals within group and cluster sized halos, to levels as small as 0.067 dex at fixed halo mass and at fixed magnitude gap. 

One obvious question is whether the measurement errors on our stellar masses are over-estimated.  Recall that our stellar masses stem from the \cite{bel03} relation, which has a 0.1 dex uncertainty at z=0.  Our BCG magnitudes also have a measurement error, which we estimate from the one-to-one comparison to the \citet{pos95} sample. These both are reasonable choices and incorporate our entire stellar mass error budget.  Just as important, our inferred intrinsic scatter when we exclude the magnitude gap in our model is $\sim 0.16$ dex, nearly identical to \citet{tin16} and \citet{kra14}.  In other words, our choice of measurement error on our stellar masses allows for a consistent comparison with those works.  When we use those same measurement errors and include the magnitude gap in our model, our intrinsic scatter drops to as low as 0.067 dex.

One challenge this presents to models like the one presented in \citet{gu2016} is that our observed intrinsic scatter is less than half of the minimum scatter allowed in their model (0.16 dex) solely from hierarchical growth.  One way to reach a smaller amount of scatter is to have a shallower SMHM relation at $z=2$.  Another option is that smooth accretion is actually not in play for BCGs (see Figures 4 and 5 in \citet{gu2016}).

In this work, we have focused on the $z \sim 0$ universe.  It is likely that the stretch factor evolves through time.  This is something that can be tested in simulations using current semi-analytic models which follow the growth history of the BCG, e.g., \citet{guo10}.  The observational challenge of an evolutionary analysis of the magnitude gap as a latent variable is to acquire good spectroscopic coverage per cluster or to understand any additional systematics which would increase the error in the magnitude gap measurement using photometric data. 

\section{Acknowledgements}
The authors would like to thank the anonymous referee for their helpful comments.  We also want to thank Juliette Becker for useful discussions about Bayesian statistics, Eric Bell for helpful discussions about the stellar mass estimates, Daniel Gifford for help with the caustic halo mass estimates, and Emmet Golden-Marx for useful discussions and for reviewing a draft of this paper. 

This material is based upon work supported by the National Science Foundation under Grant No. 1311820. The Millennium Simulation databases used in this paper and the web application providing online access to them were constructed as part of the activities of the German Astrophysical Virtual Observatory (GAVO).  Funding for SDSS-III has been provided by the Alfred P. Sloan Foundation, the Participating Institutions, the National Science Foundation, and the U.S. Department of Energy Office of Science. The SDSS-III web site is http://www.sdss3.org/.

SDSS-III is managed by the Astrophysical Research Consortium for the Participating Institutions of the SDSS-III Collaboration including the University of Arizona, the Brazilian Participation Group, Brookhaven National Laboratory, Carnegie Mellon University, University of Florida, the French Participation Group, the German Participation Group, Harvard University, the Instituto de Astrofisica de Canarias, the Michigan State/Notre Dame/JINA Participation Group, Johns Hopkins University, Lawrence Berkeley National Laboratory, Max Planck Institute for Astrophysics, Max Planck Institute for Extraterrestrial Physics, New Mexico State University, New York University, Ohio State University, Pennsylvania State University, University of Portsmouth, Princeton University, the Spanish Participation Group, University of Tokyo, University of Utah, Vanderbilt University, University of Virginia, University of Washington, and Yale University.


\end{document}